\RequirePackage{fixltx2e}
\RequirePackage{fix-cm}

\documentclass[%
floatfix,
showkeys,
twocolumn, %
nofootinbib, %
superscriptaddress, %
balancelastpage,
flushbottom,
]{revtex4-1}

\usepackage{cmap}

\usepackage{ucs}
\usepackage[utf8x]{inputenc}
\usepackage[T1,T2A]{fontenc}
\usepackage[german,russian,english]{babel}

\usepackage[sort&compress]{natbib}
\usepackage{amsmath}
\usepackage{amssymb}

\usepackage{mathtools}
\mathtoolsset{
showonlyrefs,
mathic = true
}

\allowdisplaybreaks

\usepackage{breakurl}

\usepackage{microtype}
\UseMicrotypeSet[protrusion]{alltext}

\usepackage{fixltx2e}

\usepackage{nicefrac}

\usepackage{physymb}

\makeatletter
\def\ps@pprintTitle{%
     \let\@oddhead\@empty
     \let\@evenhead\@empty
     \let\@oddfoot\@empty
     \let\@evenfoot\@oddfoot}
\makeatother
\usepackage{pgfplotstable}
\usepackage{booktabs}
\usepackage{array}
\usepackage{colortbl}

\usepackage{menukeys}

\renewcommand{\d}{\mathrm{d}}
\newcommand{\e}{\mathrm{e}}

\begin{document}
\graphicspath{{image/gns3-red/}}

\title{Designing Installations for Verification of the Model of Active Queue Management Discipline RED in the GNS3}

\author{T. R. Velieva}
\email{trvelieva@gmail.com}
\affiliation{Department of Applied Probability and Informatics\\
Peoples' Friendship University of Russia\\
Miklukho-Maklaya str. 6, Moscow, 117198, Russia}

\author{A. V. Korolkova}
\email{akorolkova@sci.pfu.edu.ru}
\affiliation{Department of Applied Probability and Informatics\\
Peoples' Friendship University of Russia\\
Miklukho-Maklaya str. 6, Moscow, 117198, Russia}

\author{D. S. Kulyabov}
\email{yamadharma@gmail.com}
\affiliation{Department of Applied Probability and Informatics\\
Peoples' Friendship University of Russia\\
Miklukho-Maklaya str. 6, Moscow, 117198, Russia}
\affiliation{Laboratory of Information Technologies\\
Joint Institute for Nuclear Research\\
Joliot-Curie 6, Dubna, Moscow region, 141980, Russia}

\thanks{Published in:
  T.~R. Velieva, A.~V. Korolkova, D.~S. Kulyabov, {Designing installations for
    verification of the model of active queue management discipline RED in the
    GNS3}, in: 6th International Congress on Ultra Modern Telecommunications and
  Control Systems and Workshops (ICUMT), IEEE, 2014, pp. 570--577.
  \href{http://dx.doi.org/10.1109/ICUMT.2014.7002164}{doi:10.1109/ICUMT.2014.7002164}.
}

\thanks{Sources:
  \url{https://bitbucket.org/yamadharma/articles-2014-gns3-red}}

\begin{abstract}
The problem of RED-module mathematical model results verification,
based on GNS3 experimental stand, is discussed in this article. The
experimental stand consists of virtual Cisco router, traffic generator
D-ITG and traffic receiver. The process of construction  of such stand
is presented. Also, the interaction between experimental stand and a
computer of investigation in order to obtain and analyze data from
stand is revised.
\end{abstract}

\maketitle

\section{Introduction}

  A stochastic model of the traffic management RED type module was
  built~\cite{korolkova:2007::en, korolkova:2009::en,
    korolkova:2010::en,
    kulyabov:2014:vestnik:red-sdu::en}. Verification of the model was
  carried out on the NS-2 basis. However, we would like to conduct
  verification on a real router. As a result was the task of designing
  an experimental stand. It was decided to verify the clean RED
  algorithm~\cite{Floyd1993} based on Cisco router. For the
  construction of the stand software package GNS3 (Graphical Network
  Simulator) was chosen.
  
  Thus, the purpose of the study is to build on the GNS3 basis
  a virtual stand consisting of a Cisco router, a traffic generator
  and a receiver. A traffic generator D-ITG (Distributed Internet
  Traffic Generator) is used as.

\section{The model of the active queue management  RED module}

  The model of the active queue management RED module is a development
  of fluid model~\cite{korolkova:2007::en,
    korolkova:2009::en, korolkova:2010::en, Misra1999, Misra2000}.  In
  the works ~\cite{kulyabov:2012:vestnik:3::en,
    kulyabov:2014:vestnik-miph:onestep::en,
    ef-kor-gev-kul-sev:vestnik-miph:2014-3} for methodology
  unification the method of one-step process randomization was
  used. Based on one-step processes we construct the stochastic model
  of RED, and the model contains two main elements: source of
  traffic and receiver.

  The source sends packets, the receiver processes the packets and
  send an acknowledge.  Our model is based on the assumption that the
  source and receiver interact with management. Thus, we obtain two
  equations, one of which describes a TCP window, and the second ---
  instant queue length. The intensity of sending packets depends on
  the window size.

  The detailed description of RED stochastic model is given in
  \cite{kulyabov:2014:vestnik:red-sdu::en} and in appendix
  (section~\ref{sec:app:construct}), where was obtained:

\begin{equation}
  \left\{
  \begin{aligned}
    \d W &= \frac{1}{T} \d t - \frac{W}{2} \d N +
    \sqrt{\frac{1}{T} + \frac{W}{2} \frac{\d N}{\d t}} \d V^1, \\
    \d Q & = \left( \frac{W}{T} - C \right) \d Q + \sqrt{\frac{W}{T}
      - C}\d V^2 ,\\
    \frac{\d \Hat{Q}}{\d t} & = w_q C ( Q - \Hat{Q} ).
  \end{aligned}
  \right.
\end{equation}
  Here $W$ is the  window size average value, $Q$ is the
  average value of the instantaneous queue, $\Hat{Q}$ is exponentially
  weighted moving average, 
  $C$ is service intensity, 
  $\d V^{i}$ is two dimensional Wiener random process.

\section{Cisco IOS  structure }

\begin{figure}
  \centering
  \includegraphics[width=0.8\linewidth]{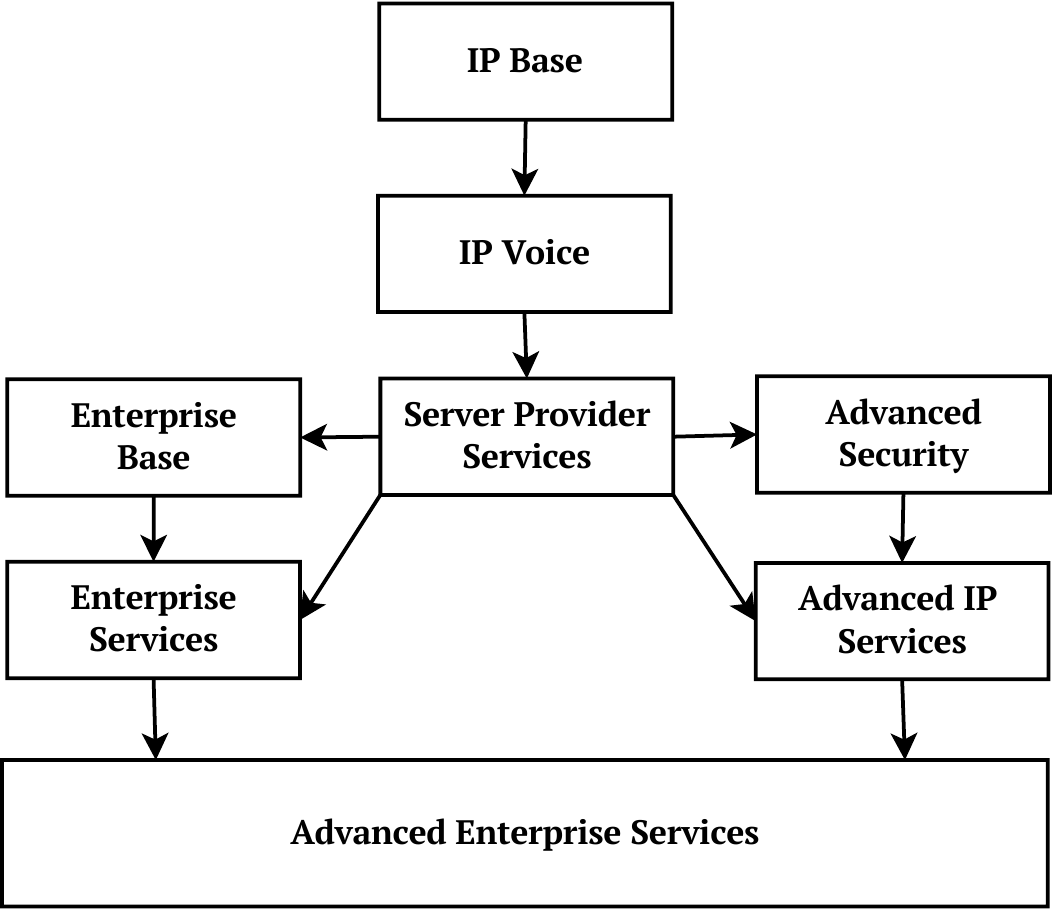}
  \caption{Feature tree of the new switch IOS naming convention}
  \label{fig:ios-tree}
\end{figure}

  To use the IOS image one should understand, 
  which features given 
  delivery option support~\cite{cisco:web:cfn}. The  parameter ``Feature
  set''~(fig.~\ref{fig:ios-tree}) is responsible for this:

  \begin{itemize}
  \item IP Base: initial level of functionality is included in all
    other sets of features. Provides a basic routing, that is static
    routes, RIP, OSPF, EIGRP, only IPv4. Includes VLAN (802.1Q and
    ISL), which previously were   available only in the IP Plus
    set. Also includes NAT.
  \item IP Services (3rd level switches): dynamic routing
    protocols, NAT, IP SLA.
  \item Advanced IP Services:  support IPv6 is added.
  \item IP Voice: functionality VoIP and VoFR is added.
  \item Advanced Security:  IOS/Firewall, IDS, SCTP, SSH and
    IPSec (DES, 3DES and AES) are added.
  \item Service Provider Services:  IPv6, Netflow, SSH, BGP,
    ATM and VoATM are added.
  \item Enterprise Base: the support for  L3 protocols
    such as IPX and AppleTalk is added. Also 
    IBM
    DLSw+, STUN/BSTUN and RSRB are added.
\end{itemize}

Since version IOS 12.3T Cisco uses a new naming scheme for 
images~(table~\ref{tab:ios-feature-new}). However, this method of 
naming does not cover all the subtleties of image acquisition, so 
elements of the old naming scheme(table~\ref{tab:ios-feature-old}) are
still used.

\begin{table}
\caption{New naming convention for Cisco Router IOS}
\label{tab:ios-feature-new}
\def\tabIosFeatureNewI{Code}
\def\tabIosFeatureNewII{Features}
\begin{center}
\pgfplotstabletypeset[col sep=&,string type,column type=l,
columns={code,feature},
columns/code/.style={column name=\tabIosFeatureNewI},
columns/feature/.style={column name=\tabIosFeatureNewII,column type=p{0.8\linewidth}},
every head row/.style={before row=\toprule,after row=\midrule},
every last row/.style={after row=\bottomrule},
]{
code &  feature
Base &  Entry level image (IP Base, Enterprise Base)
Services &      Addition of IP Telephony Service, MPLS, Voice over IP, Voice over Frame Relay and ATM (Included in SP Services, Enterprise Services)
Advanced &      Addition of VPN, Cisco IOS Firewall, 3DES encryption, SSH, Cisco IOS IPSec and Intrusion Detection Systems (IDS) (Advanced Security, Advanced IP Services)
Enterprise &    Addition of multi-protocols, including IBM, IPX, AppleTalk (Enterprise Base, Enterprise Services)
}
\end{center}
\end{table}

\begin{table}
\caption{Old naming convention for Cisco Router IOS}
\label{tab:ios-feature-old}
\begin{center}
\def\tabIosFeatureOldI{Code}
\def\tabIosFeatureOldII{Features}
\pgfplotstabletypeset[col sep=&,string type,column type=l,
columns={code,feature},
columns/code/.style={column name=\tabIosFeatureOldI},
columns/feature/.style={column name=\tabIosFeatureOldII,column type=p{0.8\linewidth}},
every head row/.style={before row=\toprule,after row=\midrule},
every last row/.style={after row=\bottomrule},
]{
code &  feature
I &     IP
Y &     IP on 1700 Series Platforms
S &     IP Plus
S6 &    IP Plus – No ATM
S7 &    IP Plus – No Voice
J &     Enterprise
O &     IOS Firewall/Intrusion Detection
K &     Cryptorgaphy/IPSEC/SSH
K8 &    56Bit DES Encryption (Weak Cryptography)
K9 &    3DES/AES Encryption (Strong Cryptography)
X &     H323
G &     Services Selection Gateway (SSG)
C &     Remote Access Server or Packet Data Serving Node (PDSN)
B &     Apple Talk
N &     Novel IP/IPX
V &     Vox
R &     IBM
U &     Unlawful Intercept
P &     Service Provider
Telco & Telecommunications Feature Set
Boot &  Boot Image (Used on high end routers/switches)
}
\end{center}
\end{table}

For unknown reasons, not all IOS images are efficient in GNS3. 
Several images were tested. Here's a short list of working and non-working images. 

Workable images:
\begin{itemize}
\item \verb|C1700-adventerprisek9-mz.124-8|.
\item \verb|C1710-bk9no3r2sy-mz.124-23|
\item \verb|C1720-l2sy-mz.121-11|
\item \verb|C2600-adventerprisek9_sna-mz.124-25b|
\item \verb|C2691-adventerprisek9_sna-mz.124-23|
\item \verb|C3660-jsx-mz.123-4.T.|
\item \verb|C3745-adventerprisek9_sna-mz.124-15.T14.|
\item \verb|C7200-adventerprisek9_sna-mz.152-4.m4|
\end{itemize}
Unworkable:
\begin{itemize}
\item \verb|C2600-adventerprisek9-sna-mz.124-23.|
\item \verb|C3745-adventerprisek9_sna-mz.124-15.T14|
\item \verb|C3745-adventerprisek9_ivs-mz.124-15.T14|
\item \verb|C3745-adventerprisek9_ivs-mz.124-15.T8|
\end{itemize}

\section{GNS3 installation and configuration}

  GNS3 allows you to simulate a virtual network consisting of routers
  and virtual machines~\cite{Welsh:2013:GNS3}. In fact, it is a
  graphical interface for different virtual machines. To emulate Cisco
  devices emulator dynamips is used. In addition, you can use such
  emulators as VirtualBox and Qemu. The latter is particularly useful
  when we work with the KVM, allowing the hardware implementation of
  the processor.  The graphical interface makes it easy to switch
  different virtual machines. Also, there is the opportunity to
  connect projected topology with the real network. Wireshark allows
  to monitor traffic within the designed topology.

\begin{figure}
   \centering
   \includegraphics[width=0.8\linewidth]{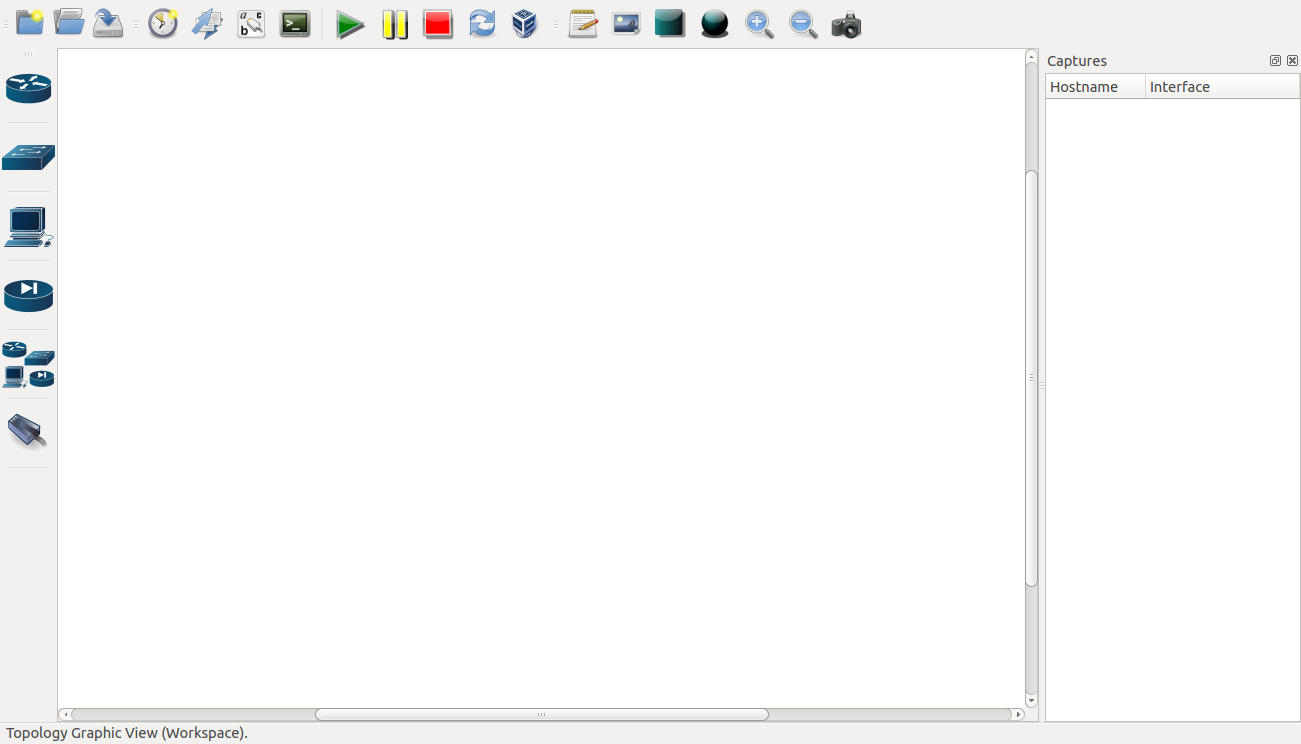}
  \caption{GNS3 interface}
\label{fig:GNS3}
\end{figure}

  To work with GNS3 we need to install Dynamips, VirtualBox and/or
  QEMU, xdotool, Wireshark. To install the above software for Linux
  operating systems (in our case, GNS3 installed on Ubuntu 14.04) the
  following commands are prescribed in the console:

\noindent \verb|sudo apt-get install dynamips|

\noindent \verb|sudo apt-get install qemu|

\noindent \verb|sudo apt-get install virtualbox|

\noindent \verb|sudo apt-get install xdotool|

\noindent \verb|sudo apt-get install wireshark|

After that  GNS3 is set with a 
command in a terminal:
\begin{verbatim}
sudo apt-get install gns3
\end{verbatim}

Then GNS3 is run. GNS3 interface opens 
(fig.~\ref{fig:GNS3}).

At the top  the context menu, on the left-hand --- the equipment to 
choose, at the bottom  
--- a console window, on the right-hand --- a network management menu. 
Before start you need to pre-configure GNS3. 

\begin{figure}
  \centering
  \includegraphics[width=0.8\linewidth]{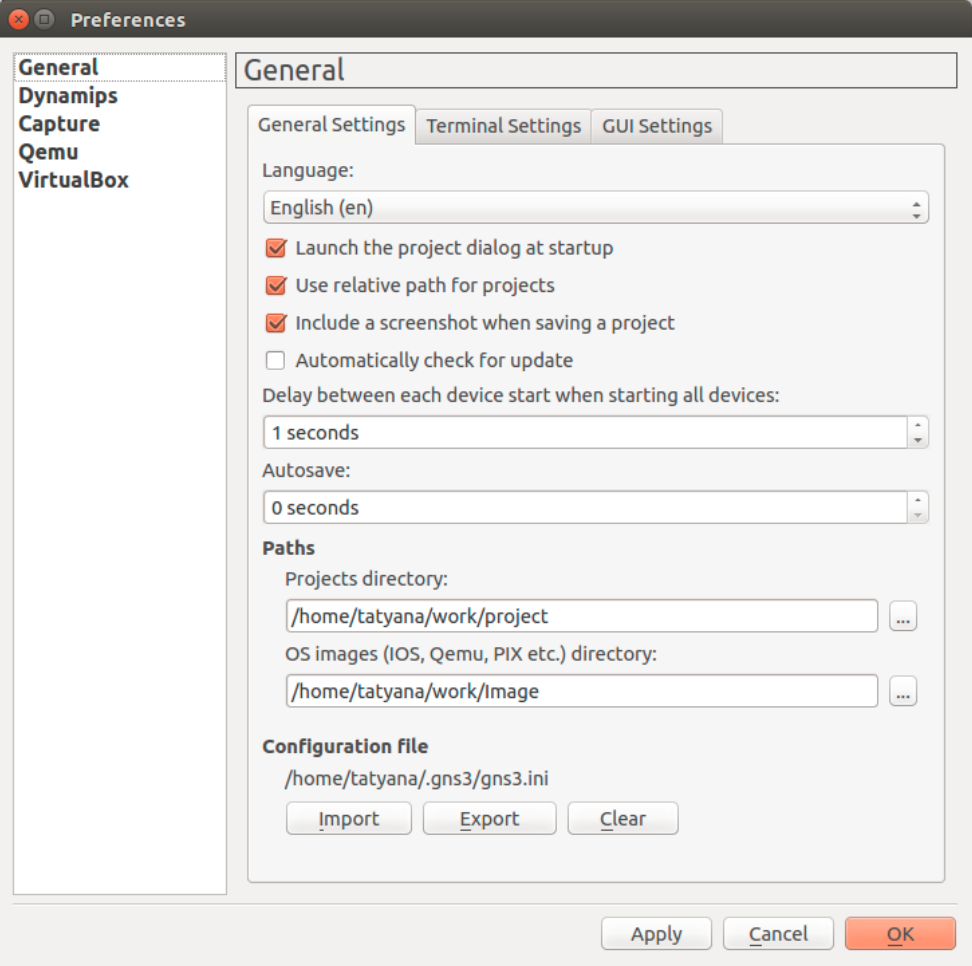}
  \caption{General preferences of GNS3}
\label{fig:Preferences:G}
\end{figure}

For this the item \menu{Preferences} is selected in 
 the \menu{Edit} of context menu. 
(fig.~\ref{fig:Preferences:G}).

\begin{figure}
  \centering
  \includegraphics[width=0.8\linewidth]{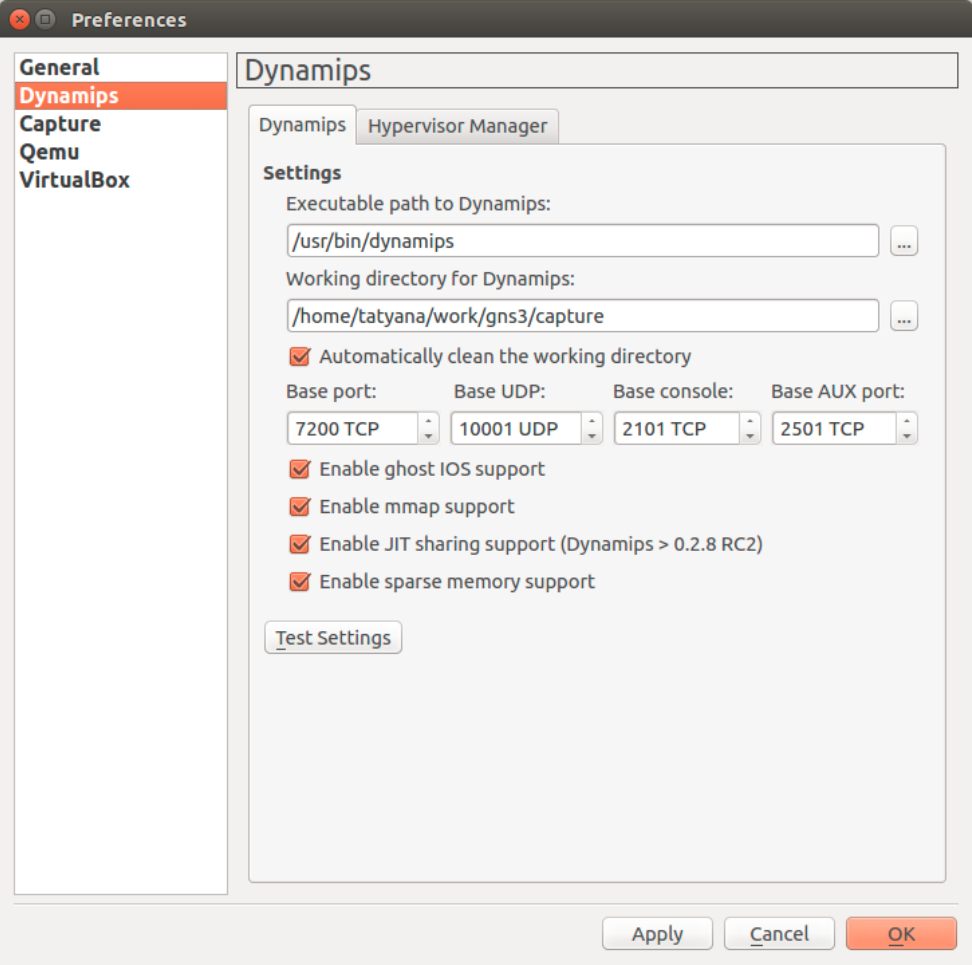}
  \caption{Dynamips preferences}
\label{fig:Preferences:D}
\end{figure}

  The language can be changed in \menu{General} submenu. Here the path
  to the folder with designs and images of equipment is prescribed
  (fig.~\ref{fig:Preferences:D}).

  In  submenu \menu{Capture} traffic capture parameters are configured
  (fig.~\ref{fig:Preferences:C}) (however, in this article 
  we do
  not use the opportunity to capture the traffic).  In line Default
  Presents the option  Wireshark Live Traffic Capture (Linux) is choosed. In the next
  line,  the path to the directory  to store data
  capture is specified. And the last line the command to start Wireshark
  capture is proscribed:
\begin{verbatim}
tail -f -c +0b %c | wireshark -k -i
\end{verbatim}
  
\begin{figure}
  \centering
  \includegraphics[width=0.8\linewidth]{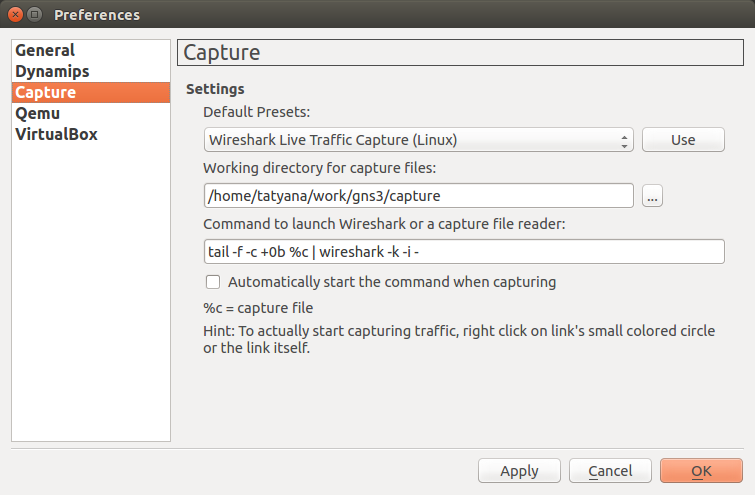}
  \caption{Capture preferences}
\label{fig:Preferences:C}
\end{figure}

\begin{figure}
   \centering
   \includegraphics[width=0.8\linewidth]{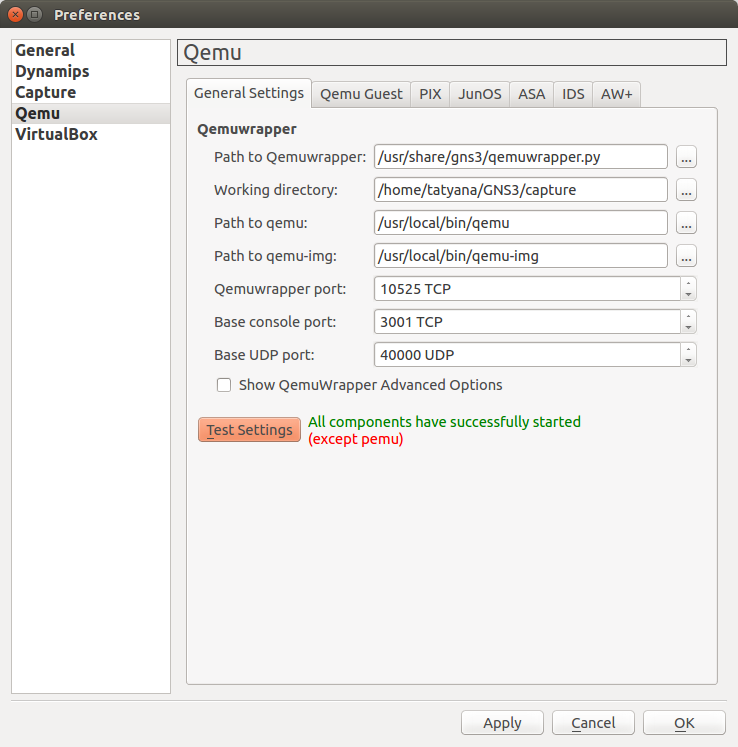}
   \caption{Preferences GNS3 - Qemu}
   \label{fig:Preferences:Qemu:g}
\end{figure}

  In the \menu{Qemu} the tab \menu{General Settings}  sets the path
 to  Qemuwrapper (this file is supplied with GNS3).  The directory for
  capture is specified. In line \menu{Path to qemu}  the path to the file that
  emulates the processor is set. The following line specifies the path to the
  virtual machine.  Port numbers are reserved  by default. To test
  Qemuwrapper one need to click on the button \menu{Test Settings},  a
  green label on the implementation of the test should appear
  (fig.~\ref{fig:Preferences:Qemu:g}).

  Next we need to install the OS image to the virtual machine. Set of images can
  be taken from the page \url{http://www.gns3.net/appliances/}. We
  chose Linux Core 4.7.7, because it contains a traffic generator
  D-ITG.

\begin{figure}
  \centering
  \includegraphics[width=0.8\linewidth]{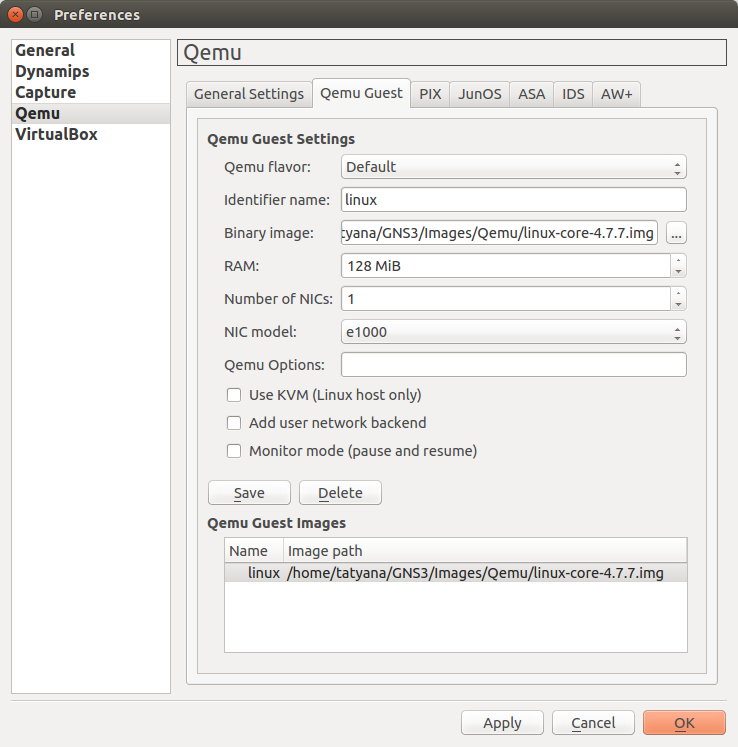}
  \caption{Preferences GNS3}
  \label{fig:Preferences:Qemu:G}
\end{figure}

  In line \menu{Qemu flavor} we specify the model and name of
  processor. 
  After that  the path to the OS
  image is set.  Push \menu{Save} and \menu{OK}. After that, the 
  \menu{Qemu guest} can be selected on the toolbox (fig.~\ref{fig:Preferences:Qemu:G}).

\begin{figure}
   \centering
   \includegraphics[width=0.8\linewidth]{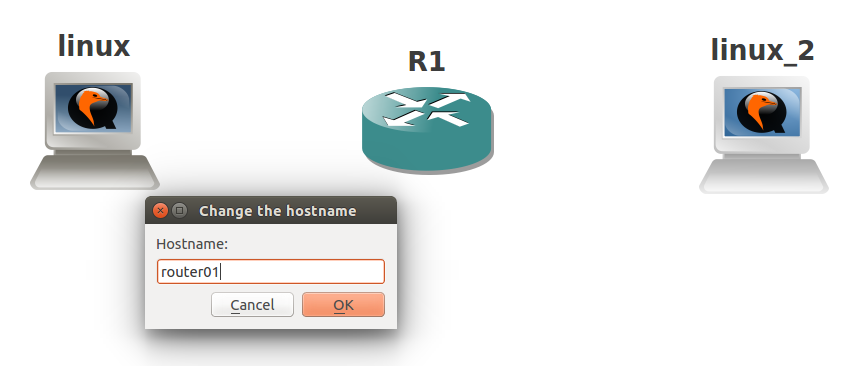}
   \caption{Virtual Network}
   \label{fig:Stand}
\end{figure}

Now you can assemble the stand. For this we portable from device selection
  menu router and virtual machine \menu{Qemu guest}, rename
  (\menu{right mouse button> change the hostname})
  (fig.~\ref{fig:Stand}).

\begin{figure}
   \centering
   \includegraphics[width=0.8\linewidth]{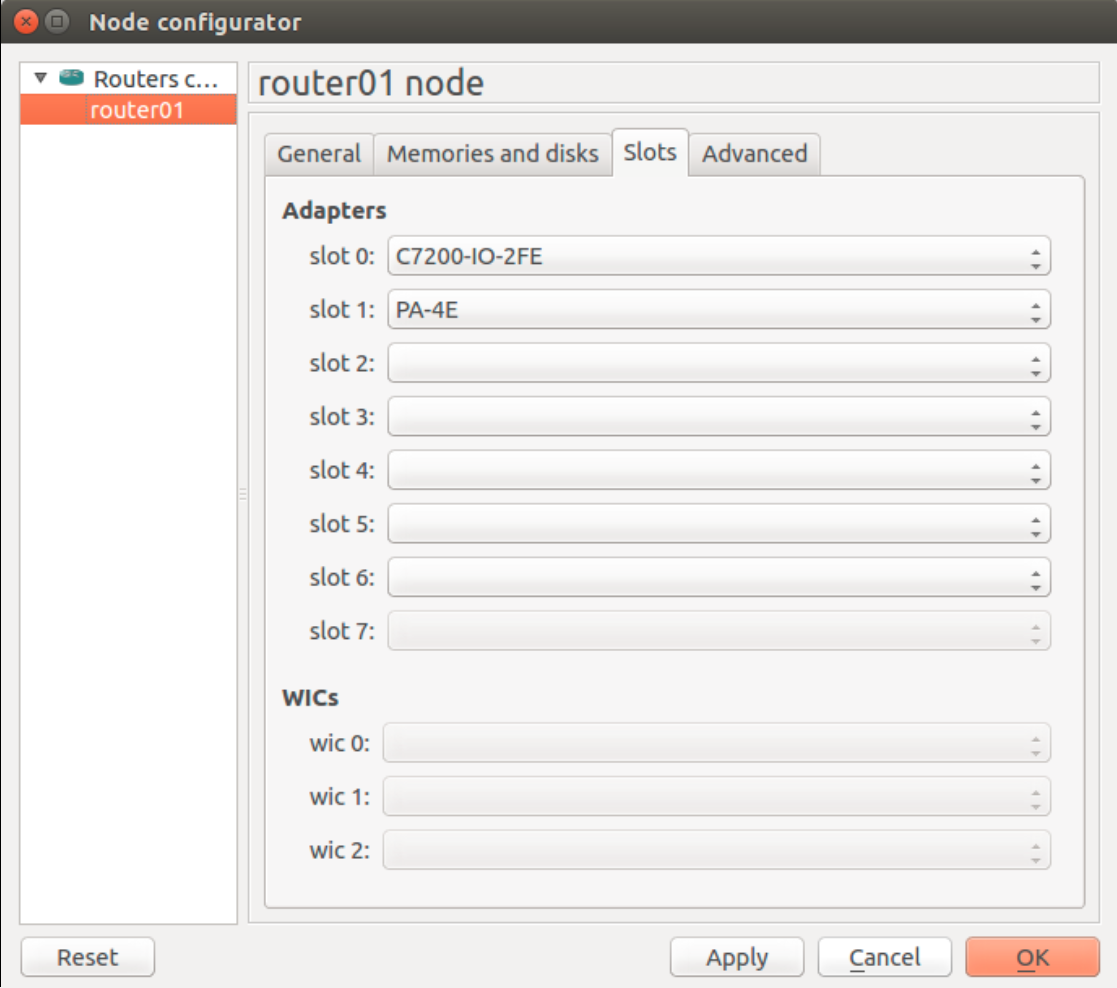}
   \caption{Router configurator}
   \label{fig:Node:R1}
 \end{figure}

  Host01 is a package source; host02 is a
  destination. Let's configure the router slots. To do this, right click on
   router01 and select  \menu{Configure> router01> Slots}.
 For slot0 we select FastEthernet (fig.~\ref{fig:Node:R1}).

  Then we create a connection between the router and the virtual machine.
  We should right-click on host01, select \menu{add link > FastEthernet > e0} and
  connect with router01, choosing f0/0. 
  A
  connection between host02 and router01 is created similarly, only
  now we choose  Ethernet type of
  connection (speed 10 Mbps).

\begin{figure}
  \centering
  \includegraphics[width=0.8\linewidth]{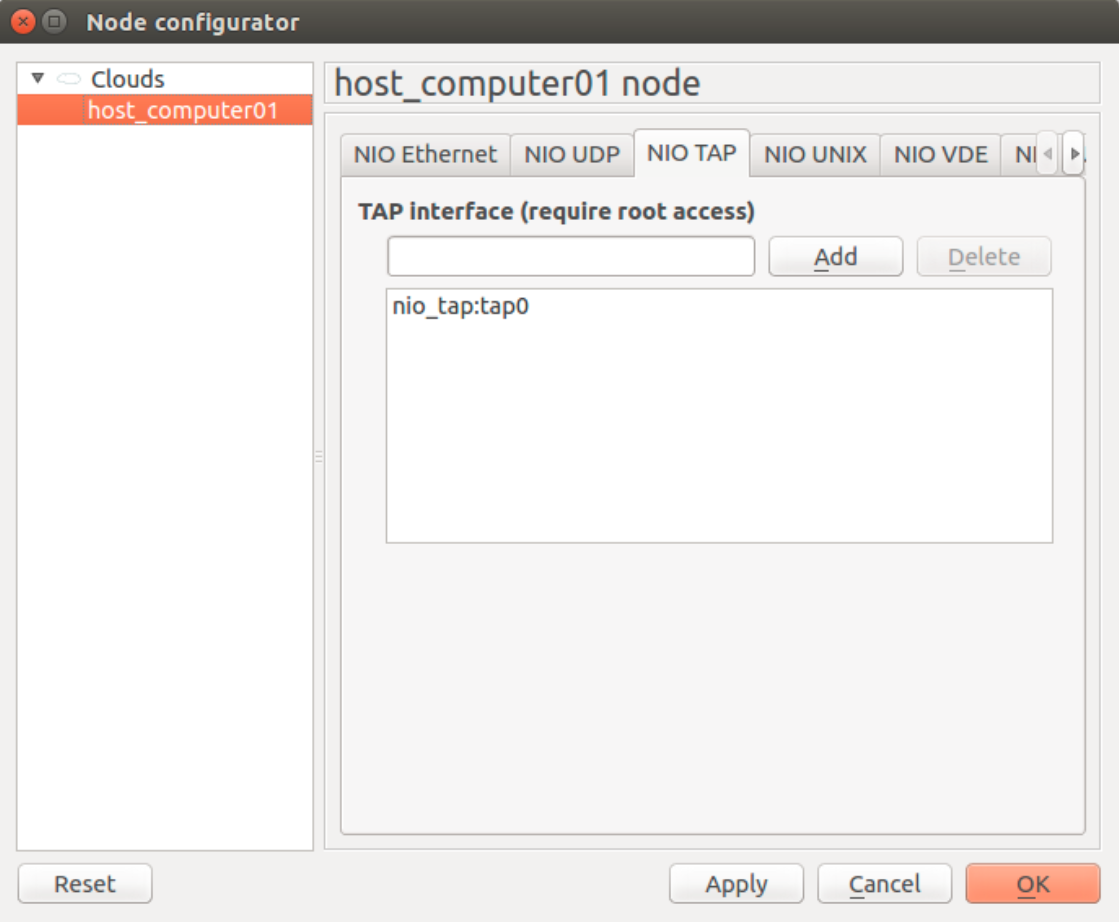}
  \caption{Cloud interface configurator}
  \label{fig:NIO:TAP}
\end{figure}

  Now we need to connect the virtual machines with the host computer
  (for receiving and processing the logs). Connection will be carried
  through TUN/ TAP interface (TAP simulates an Ethernet device and
  works at the link layer model OSI, Ethernet frames and terms used to
  create a bridge). To do this, drag it to the workspace cloud and
  configure it: \menu{Configure> NIO TAP}.  Write the name tap0 and press
  \menu{add> apply> ok} (fig.~\ref{fig:NIO:TAP}).

\begin{figure}
  \centering
  \includegraphics[width=0.8\linewidth]{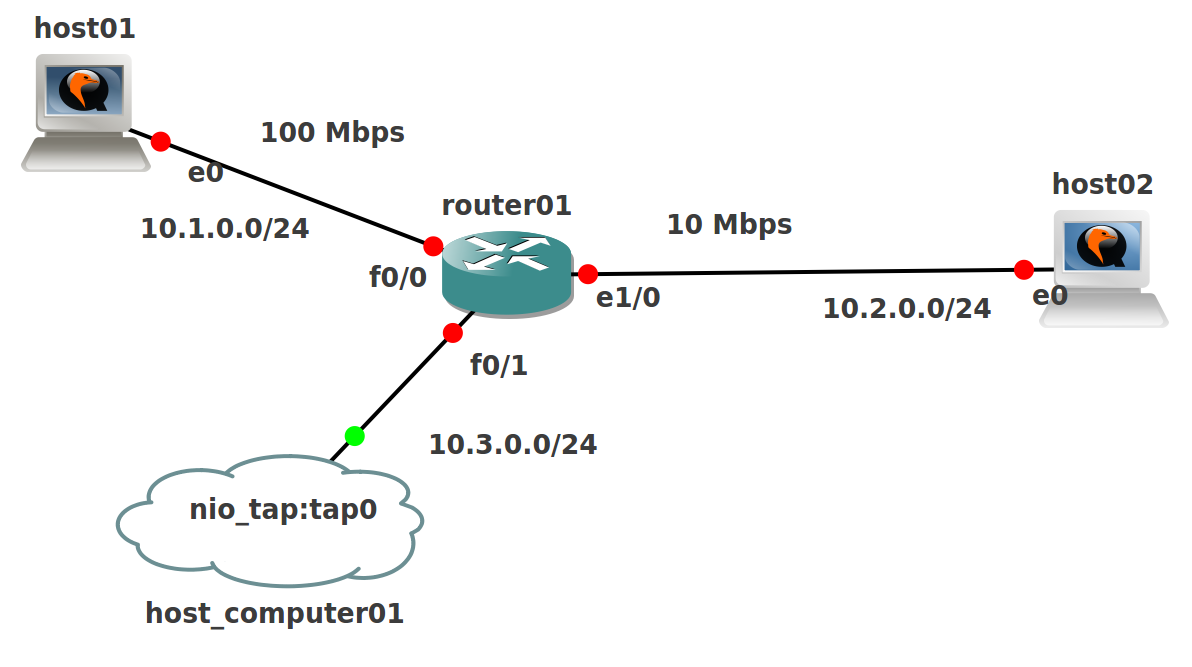}
  \caption{Virtual Network}
  \label{fig:VN}
\end{figure}

We create a connection to router01, selecting the type of connection 
FastEthernet, and get topology (fig.~\ref{fig:VN}). 

Next, you need to configure the devices.

\begin{figure}
  \centering
  \includegraphics[width=0.8\linewidth]{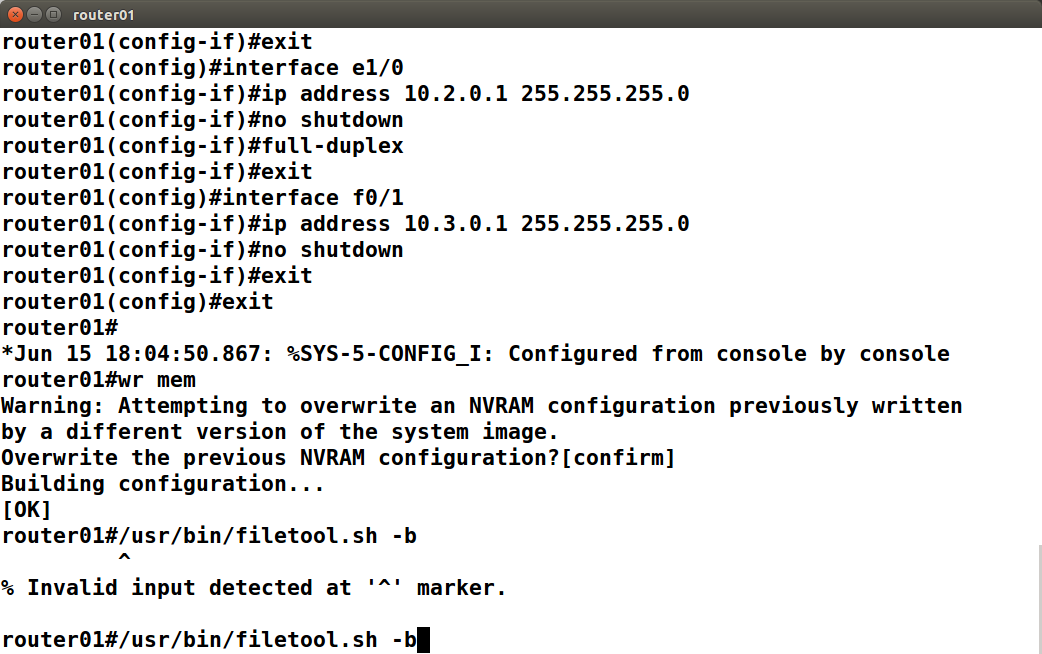}
  \caption{Router CLI configuration}
  \label{fig:router}
\end{figure}

  Run router01 by clicking (\menu{right button > start}). Open the console for it
  (\menu{right button > Console}) and enter the following commands
  (fig.~\ref{fig:router}):
\begin{verbatim}
Router>enable
Router#configure terminal
router01(config)#hostname router01
router01(config)#interface f0/0
router01(config-if)#ip address 10.1.0.1 
  255.255.255.0
router01(config-if)#no shutdown
router01(config-if)#exit
router01(config)#interface e1/0
router01(config-if)#ip address 10.2.0.1 
  255.255.255.0
router01(config-if)#no shutdown
router01(config-if)#full-duplex
router01(config-if)#exit
router01(config)#interface f0/1
router01(config-if)#ip address 10.3.0.1 
  255.255.255.0
router01(config-if)#no shutdown
router01(config-if)#exit
router01(config)#exit
router01#write memory
\end{verbatim}

  Now we open the console for host01. As the name of the user we specifie
  \verb|tc|, the password is blank. Go to the superuser:
  \verb|sudo su|.  Edit parameters host01 (to start editing in vi we
  should 
  enter \verb|i|; to save the changes and exit we should  press the
  button esc, then enter \verb|:wq| and press the \keys{Enter}). We need to edit
  two files: \verb|/opt/bootsync.sh| and \verb|/opt/bootlocal.sh|:
\begin{verbatim}
vi /opt/bootlocal.sh
\end{verbatim}

At end of file we write the following:
\begin{verbatim}
ifconfig eth0 10.1.0.10 netmask 255.255.255.0 up
route add default gw 10.1.0.1
\end{verbatim}

  After  saving the changes to a file we exit from it, and execute the
  command to save the configuration (Linux Core uses the  Tinycore 
  configuration system):
\begin{verbatim}
/usr/bin/filetool.sh -b
\end{verbatim}

Similarly host02 is configured, but ip address (network 10.2.0.0/24)
should be changed.

Let's configure the host  computer with TAP-interface. 
In the command window as a root we  create an interface for  user:
\begin{verbatim}
tunctl -u user_name -t tap0
\end{verbatim}

Now we set the address for tap0 interface (fig.~\ref{fig:Tap0:1}):
\begin{verbatim}
ifconfig tap0 10.3.0.10 netmask 255.255.255.0 up
\end{verbatim}

\begin{figure}
  \centering
  \includegraphics[width=0.8\linewidth]{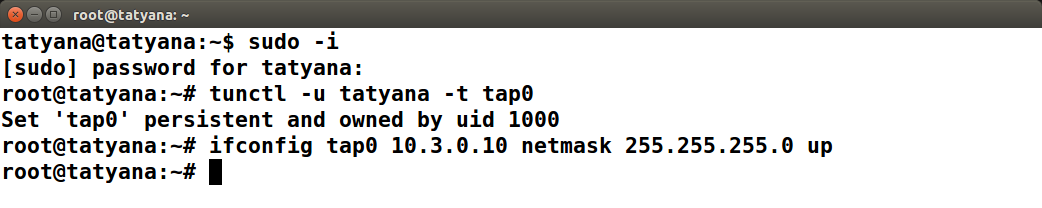}
  \caption{Address configuration on tap0 interface}
  \label{fig:Tap0:1}
\end{figure}

Then you can  configure filtering and routing (fig.~\ref{fig:Tap0}):
\begin{verbatim}
iptables -I INPUT 1 -i tap0 -j ACCEPT
route add -net 10.1.0.0/24 dev tap0
route add -net 10.2.0.0/24 dev tap0
\end{verbatim}

\begin{figure}
  \centering
  \includegraphics[width=0.8\linewidth]{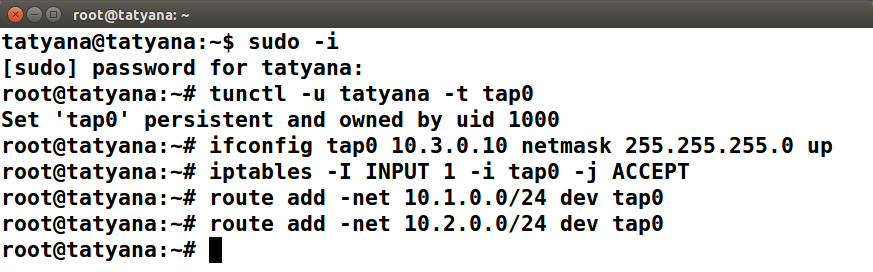}
  \caption{Routing configuration on tap0 interface}
  \label{fig:Tap0}
\end{figure}

\section{Traffic Generator D-ITG}

Now we can generate traffic and take readings. To generate traffic we use D-ITG.

D-ITG provides estimates for key indicators of quality of service (the 
average packet delay, delay variation (jitter), packet loss ratio, 
throughput) with a high degree of certainty. Depending on the 
requirements 
of the 
experiment, we can change 
the following parameters:

\begin{itemize}
\item number of streams transmitted between end stations;
\item duration of traffic generation;
\item the intensity of each separate stream (pack/s or bit/s);
\item packet traffic length  or distribution law;
\item type and parameters of the distribution law of the time interval 
between adjacent packets (for example, packet length and time interval 
between packets can be distributed in a uniform, exponential, normal, 
gamma-law or Pareto, Cauchy, Poisson);  
\item type of transport layer protocols: TCP, UDP, DCCP, SCTP;
\item type of traffic (simulation of a flow generated by specific application 
protocol): VoIP, IPTV, Telnet, DNS, game traffic (Counter Strike, 
Quake3).  
\end{itemize}

Within the D-ITG the control of experiment is performed by using the command line, 
and the required set of parameters to generate traffic is given by 
calling the program \verb|ITGSend| using keys.

\begin{figure}
   \centering
   \includegraphics[width=0.8\linewidth]{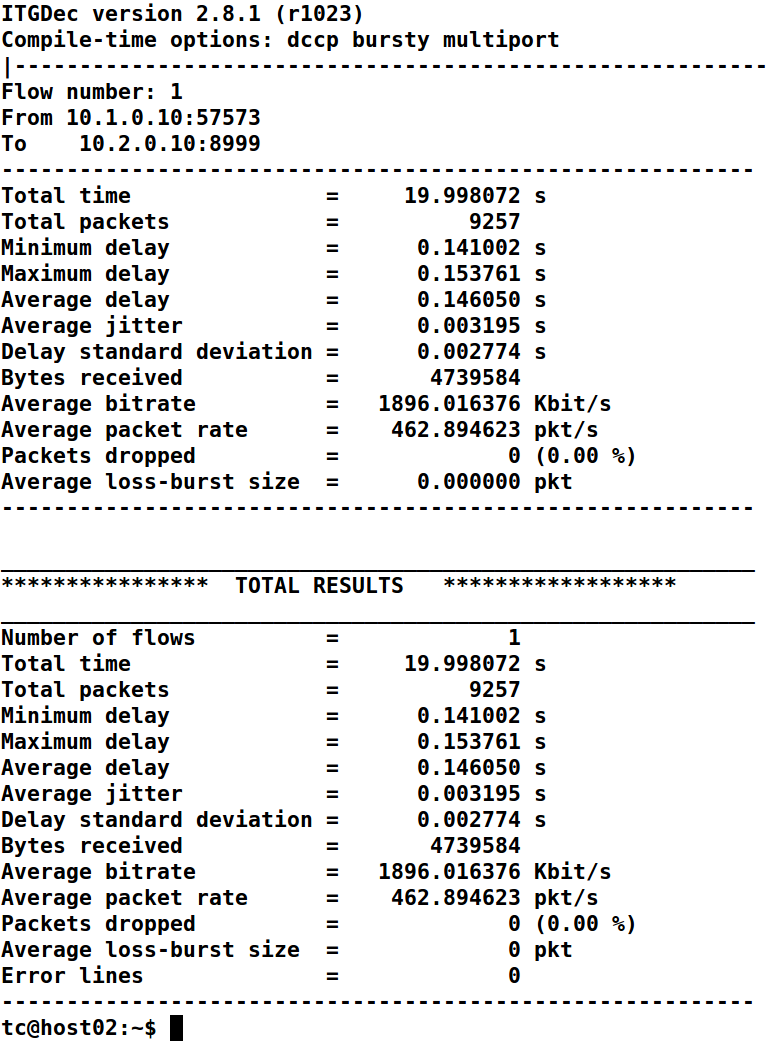}
  \caption{Input flow statistics. UDP}
\label{fig:UDP}
\end{figure}

Let's consider the example of generating traffic for UDP and TCP.
In host02 console we run the command  of channel audition:
\begin{verbatim}
ITGRecv
\end{verbatim}

In host01 console we run host01 traffic generation:
\begin{verbatim}
ITGSend -a 10.2.0.10 -T UDP -C 10000 -c 500 \
  -t 20000 -x recv_log_file
\end{verbatim}

Here we transfer the files to the receiver address by 10.2.0.10 protocol
UDP, the transmission rate is  10,000 packets per second, the size of
package is 500
bytes, the connection time is 20000 ms. All the information about
send data is  recorded on the receiving end in a file called
\verb|recv_log_files|. Once data transfer has been performed,
we stop listening to host02  console  (pressing \keys{Ctrl+C}), then
run the command in host02 console  to decode the log file:
\begin{verbatim}
ITGDec recv_log_file
\end{verbatim}

By this command  a table with the values ​​of the received stream is
displayed (fig.~\ref{fig:UDP}).

\begin{figure}
  \centering
  \includegraphics[width=0.8\linewidth]{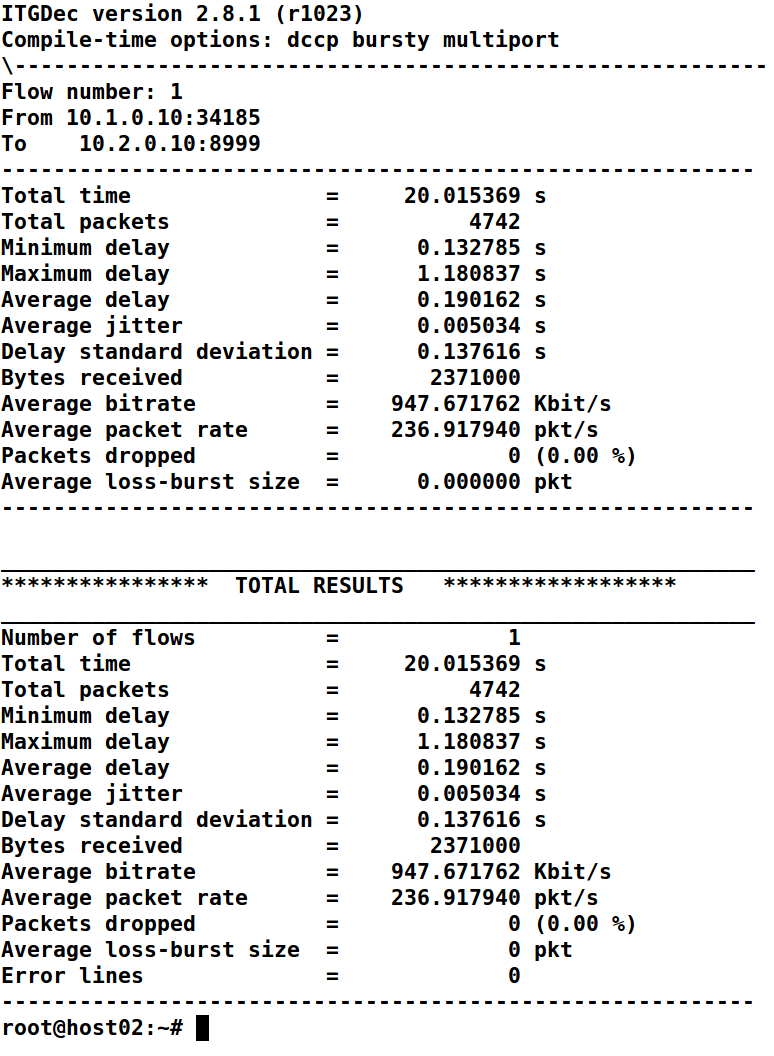}
  \caption{Input flow statistics. TCP}
  \label{fig:TCP}
\end{figure}

We take similarly steps for TCP and get  the following results (fig.~\ref{fig:TCP}).

\begin{figure}
  \centering
  \includegraphics[width=0.8\linewidth]{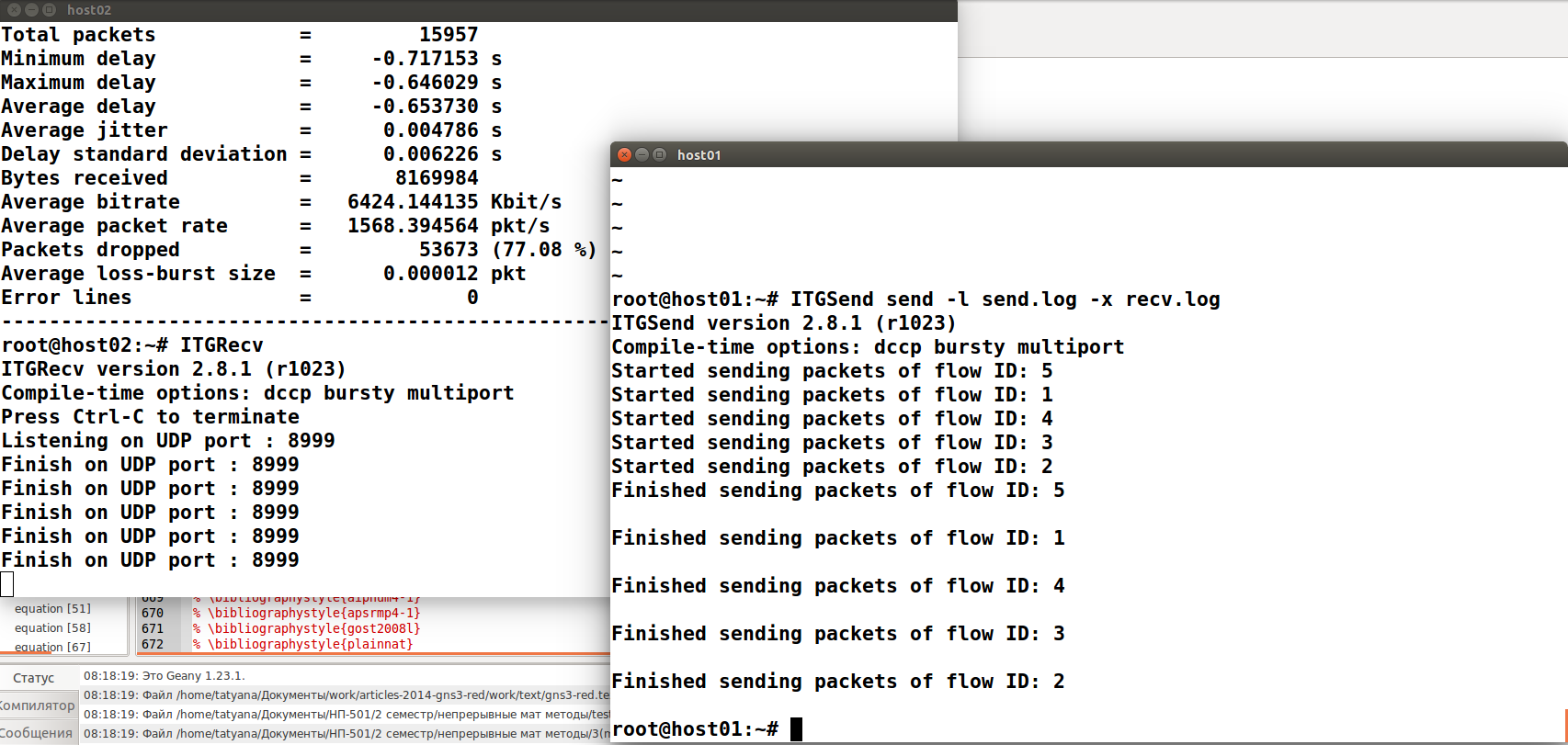}
  \caption{Log decryptiion}
\label{fig:Command:UDP}
\end{figure}

  For multicast traffic in host01 command window for file \verb|send|
   we describe each flow of traffic. For example, as follows:

\begin{verbatim}
cat > send <<EOF
-a 10.2.0.10  -C 1000 -c 512 -T UDP
-a 10.2.0.10  -C 2000 -c 512 -T UDP
-a 10.2.0.10  -C 3000 -c 512 -T UDP
-a 10.2.0.10  -C 4000 -c 512 -T UDP
-a 10.2.0.10  -C 5000 -c 512 -T UDP
EOF
\end{verbatim}

  Now let's run the traffic flow from host01.  The transmission
  traffic data from
  host01 to host02 is written in the files  send.log and recv.log:
\begin{verbatim}
ITGSend send -l send.log -x recv.log
\end{verbatim}

Let's display the report of flow transmission. To do this, we write the command console host02
\begin{verbatim}
ITGDec send.log
\end{verbatim}
to decrypt the sent traffic  (fig.~\ref{fig:Command:UDP}) and
command
\begin{verbatim}
ITGDec recv.log
\end{verbatim}
to decrypt the received traffic.

  Thus, we obtain data about each flow, as well as the value of the
  outgoing and incoming flows  parameters.

\section{Visualization of results}

Charts can be constructed for the following parameters:
 delay,
 jitter,
 bitrate,
packet loss. 

\begin{figure}
  \centering
  \includegraphics[width=0.8\linewidth]{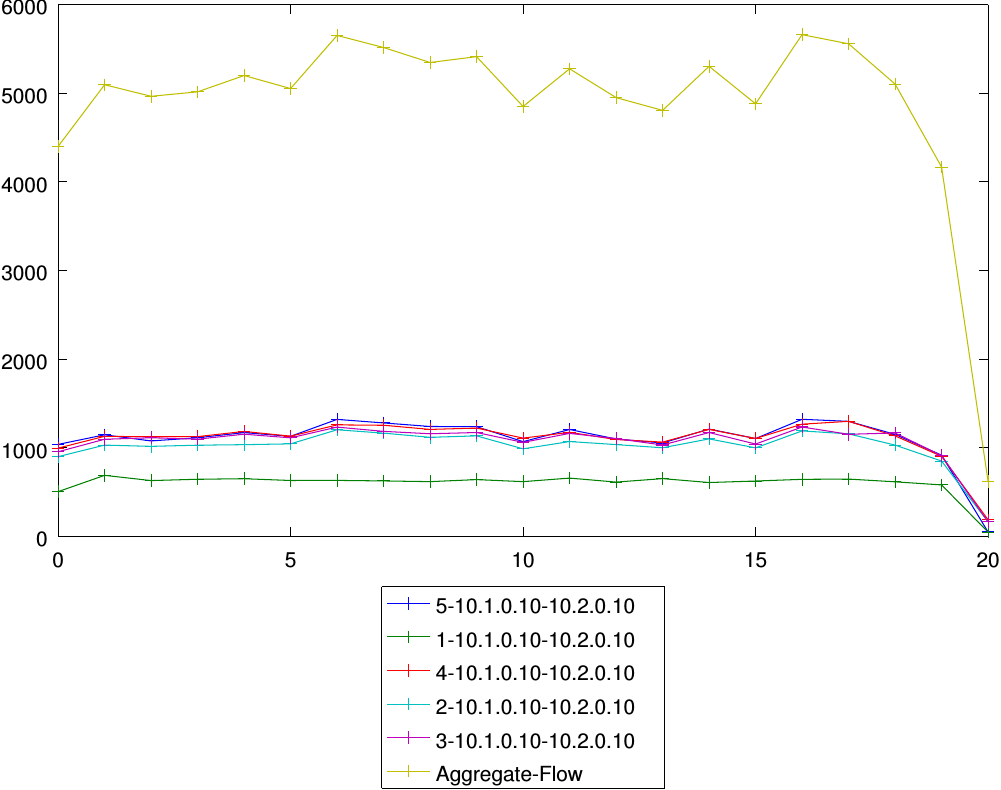}
  \caption{UDP packets loss}
\label{fig:Command:UDP:Packetloss}
\end{figure}

  Assume that we have 5 streams,
  which are transmitted over TCP. Flow rate 1 is 1000 pps, flow 2 is
  2000 pps, stream 3 is 3000 pps, stream 4 is 4000 pps, stream 5 is 5000
  pps. Packet sizes are 512 bytes and the same for all streams, and
  Connection Time is 20000 ms. For packet loss demonstration we will
  use UDP protocol 
  (fig.~\ref {fig:Command:UDP:Packetloss}).

\begin{figure}
   \centering
   \includegraphics[width=0.8\linewidth]{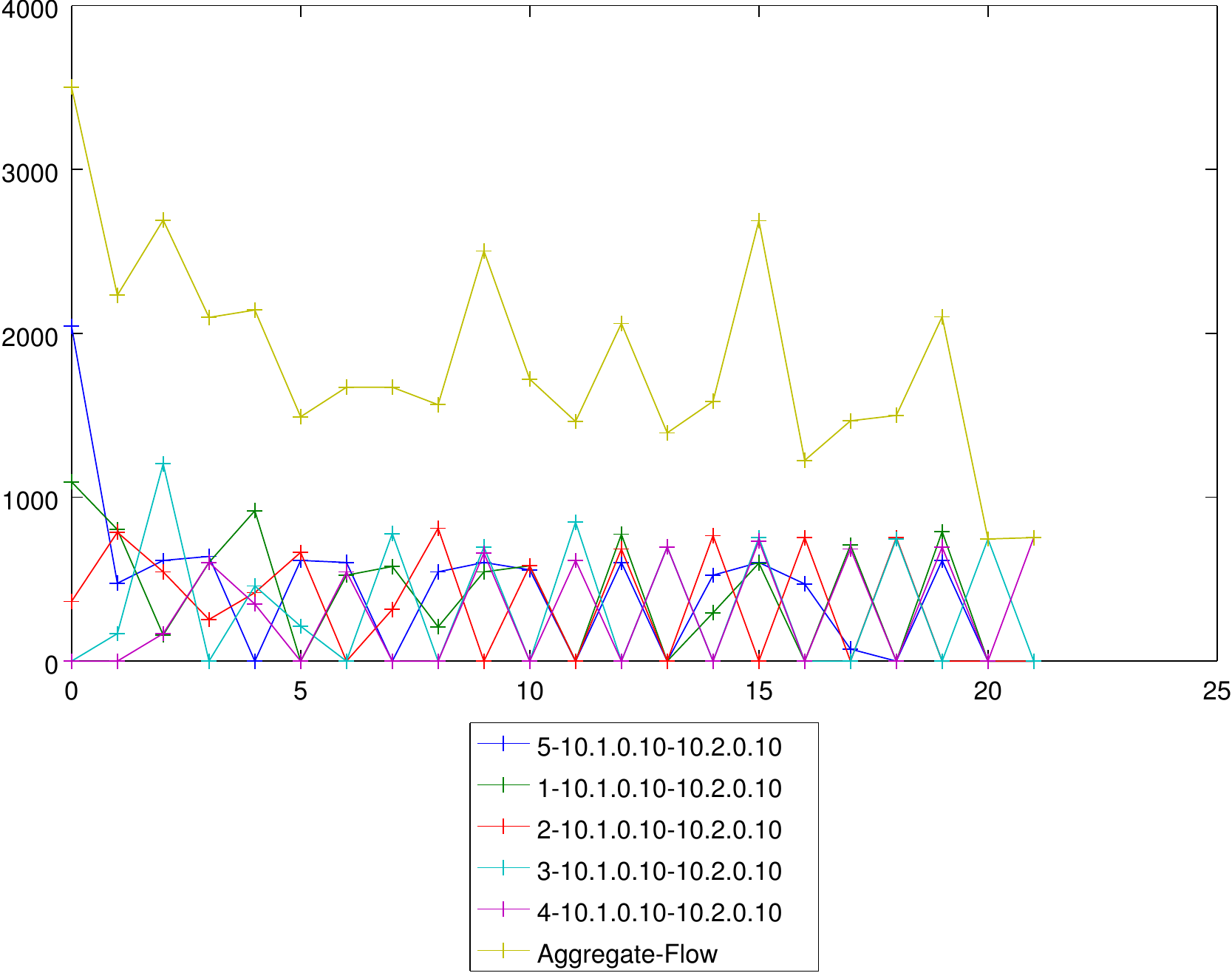}
  \caption{Incoming bitrate}
 \label{fig:Command:TCP:Bitrate:host01}
\end{figure}

\begin{figure}
  \centering
  \includegraphics[width=0.8\linewidth]{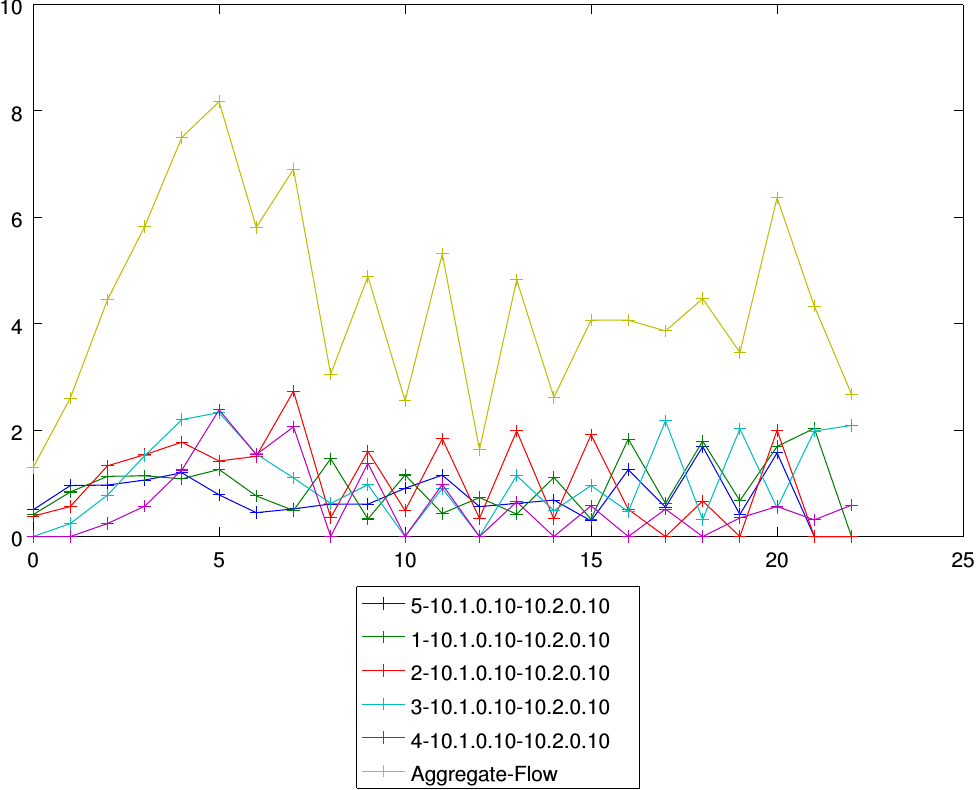}
  \caption{Incoming delay}
\label{fig:Command:TCP:Delay:host02}
\end{figure}

\begin{figure}
  \centering
  \includegraphics[width=0.8\linewidth]{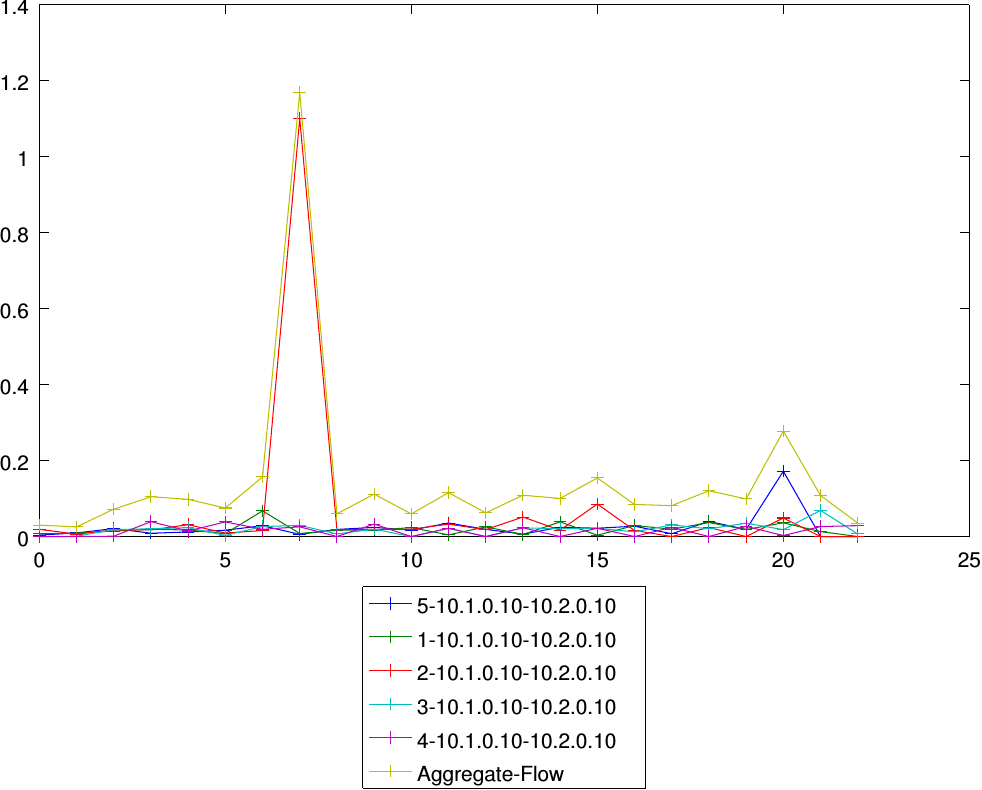}
  \caption{Incoming jitter}
\label{fig:Command:TCP:jitter}
\end{figure}

With the help of \verb |ITGplot| we make the graphs of bitrate, delay and jitter. For this
we need to record the values ​​obtained in individual files using the following commands:
\begin{verbatim}
ITGDec recv.log -b 1000 
ITGDec recv.log -d 1000
ITGDec recv.log -j 1000  
\end{verbatim}

 Every 1000 ms in files bitrate.dat, jitter.dat, delay.dat
the respective values ​​of the parameters are recorded for the transmitted flows
(fig.~\ref{fig:Command:TCP:Bitrate:host01},
\ref{fig:Command:TCP:Delay:host02}, \ref{fig:Command:TCP:jitter}).
We construct graphs using the following commands:
\begin{verbatim}
./ITGplot birate.dat
./ITGplot delay.dat
./ITGplot jitter.dat
\end{verbatim}

On obtained graphs  the upper curve shows the general behavior of all flows.

\section{Conclusion}

Thus, we have built the foundation of the stand for  verification model
of traffic management module RED. However, in this paper is not considered
However, the task of RED module configuration on given router is not
considered as well as a problem of real rime router reading.

\appendix*

\section{Active queue management  RED module stochastic model construction }
\label{sec:app:construct}

The window change for each acknowledgement:
\begin{equation}
  W_{n+1} = W_n + \frac {1}{W_n}.
\end{equation}

For one-step processes stochastization we need to use the continuous time model. 
In round-trip time $T$ there will be $W_{n}$ acknowledgments:
\begin{equation}
  \Dot{W} = \frac{1}{T}.
\end{equation}

The discrete equation of  queue instantaneous length:
\begin{equation}
  Q_{n+1} = Q_{n} + W_{n} - C_{n}.
\end{equation}
 
Based on the discrete equation of 
 queue instantaneous length the following equation is derived:
\begin{equation}
  \Dot{Q} = \frac{W(t)}{T(t)} - C(t).
\end{equation}

Next, the behaviour of the exponentially weighted moving average 
queue length (the equation of connection between the source 
and the recipient) will be described.

The discrete recurrent equation of average queue length:
\begin{equation}
  \Hat{Q}(t_{k} + \delta) = (1 - w_q)\Hat{Q}(t_k) +w_q Q(t_k).
\end{equation}
From above mentioned discrete equation the continuous equation of average queue length 
is derived:
\begin{equation}
  \Dot{\Hat{Q}}= - \frac{w_q}{\delta} \Hat{Q} + \frac{w_q}{\delta} Q.
\end{equation}

In order to consider the packets behavior  in the system the kinetic 
equation, or as it is called the equation of interaction, will be presented. 
The number of packets is specified by the TCP window .
\begin{equation}
  \begin{cases}
    0 \xrightarrow{k^1_1} W,\\
    W \xrightarrow{k^1_2} 0.
  \end{cases}
\end{equation}
  The first relation describes the appearance of packages in the
  system, the second describe departure of the packages from the
  system.

Based on the written equations and with the help of the method of constructing 
one-step processes the Fokker--Planck equation is derived:
\begin{multline}
  \frac{\partial w}{\partial t} = 
  - \frac{\partial}{\partial W} 
  \left[ \left( \frac{1}{W} - \frac{W}{2} \frac{\d N}{\d t} \right) w \right]
  + {} \\ {} +
  \frac{1}{2} \frac{\partial^2}{\partial W^2} 
  \left[ \left( \frac{1}{W} + \frac{W}{2} \frac{\d N}{\d t} \right) w \right].
\end{multline}
From the Fokker--Planck equation the corresponding 
Langevin  equation is obtained.
\begin{equation}
  \label{eq:W:langevin}
  \d W = \frac{1}{W} \d t - \frac{W}{2} \d N +
  \sqrt{\frac{1}{W} + \frac{W}{2} \frac{\d N}{\d t}} \d V^1,
\end{equation}
where $\d V^1$ is  the Wiener process corresponding to the random 
process $W(t)$.

Similarly, the behaviour of the queue length is described and interaction equations 
for the instantaneous queue length are given:
\begin{equation}
  \begin{cases}
    0 \xrightarrow{k^2_1} Q\\
    0\xrightarrow{k^2_2} Q.
  \end{cases}
\end{equation}

The Fokker--Planck equation for the instantaneous queue length:
\begin{equation}
  \frac{\partial q}{\partial t} = 
  - \frac{\partial}{\partial Q} 
  \left[ \left( \frac{W}{T} - C \right) q \right]
  + \frac{1}{2} \frac{\partial^2}{\partial Q^2} 
  \left[ \left( \frac{W}{T} - C \right) q \right].
\end{equation}

The Langevin equation for the instantaneous queue length:
\begin{equation}
  \label{eq:Q:langevin}
  \d Q = \left( \frac{W}{T} - C \right) \d t +
  \sqrt{\frac{W}{T} - C} \d V^2,
\end{equation}
where $\d V^2$ is the Wiener process which corresponds to the random 
process  $Q$.

According to the obtained equations the resulting system of the equations is given:
\begin{equation}
  \left\{
  \begin{aligned}
    \d W &= \frac{1}{T} \d t - \frac{W}{2} \d N +
    \sqrt{\frac{1}{T} + \frac{W}{2} \frac{\d N}{\d t}} \d V^1, \\
    \d Q & = \left( \frac{W}{T} - C \right) \d Q + \sqrt{\frac{W}{T}
      - C}\d V^2 ,\\
    \frac{\d \Hat{Q}}{\d t} & = w_q C ( Q - \Hat{Q} ).
  \end{aligned}
  \right.
\end{equation}
The detailed stochastic model of the router RED-like control module 
 is described in~\cite{kulyabov:2014:vestnik:red-sdu::en}.

\bibliographystyle{elsarticle-num}

\bibliography{bib/gns3-red,bib/gns3-red_ru,bib/gns3-red_en}

\end{document}


\graphicspath{{image/gns3-red/}}

\title{Конструирование установки для верификации модели модуля активного управления трафиком RED в среде GNS3}

\author{Т. Р. Велиева}
\email{trvelieva@gmail.com}
\affiliation{Кафедра прикладной информатики и теории вероятностей,\\
  Российский университет дружбы народов,\\
  ул. Миклухо-Маклая, д.6, Москва, Россия, 117198}

\author{А. В. Королькова}
\email{akorolkova@sci.pfu.edu.ru}
\affiliation{Кафедра прикладной информатики и теории вероятностей,\\
  Российский университет дружбы народов,\\
  ул. Миклухо-Маклая, д.6, Москва, Россия, 117198}

\author{Д. С. Кулябов}
\email{yamadharma@gmail.com}
\affiliation{Кафедра прикладной информатики и теории вероятностей,\\
  Российский университет дружбы народов,\\
  ул. Миклухо-Маклая, д.6, Москва, Россия, 117198}
\affiliation{Лаборатория информационных технологий,\\
Объединённый институт ядерных исследований,\\
ул. Жолио-Кюри 6, Дубна, Московская область, Россия, 141980}

\thanks{Опубликовано в: 
\emph{Velieva~T.~R., Korolkova~A.~V., Kulyabov~D.~S.}
  Designing installations
  for verification of the model of active queue management discipline RED in
  the GNS3~// 6th International Congress on Ultra Modern Telecommunications
  and Control Systems and Workshops (ICUMT).~---
  IEEE, 2014.~---
  P.~570--577.~---
  \href{http://dx.doi.org/10.1109/ICUMT.2014.7002164}{doi:10.1109/ICUMT.2014.7002164}.
}

\thanks{Исходные тексты:
  \url{https://bitbucket.org/yamadharma/articles-2014-gns3-red}}

\begin{abstract}
При построении стохастической модели RED возникла проблема проведения
этапа верификации. В качестве системы верификации предложено
использовать реализацию протокола RED компании Cisco. Установка
собрана в симуляторе оборудования Cisco — системе GNS3. 
\end{abstract}

\maketitle

\section{Введение}

  В нашем коллективе была построена стохастическая модель модуля
  управления трафиком типа RED~\cite{korolkova:2007, korolkova:2009,
    korolkova:2010, kulyabov:2014:vestnik:red-sdu}. Верификация модели
  была проведена на основе средства имитационного моделирования
  NS-2. Однако нам хотелось бы провести верификацию на реальном
  маршрутизаторе. В результате возникла задача по конструированию
  экспериментального стенда. Для начала было решено верифицировать
  чистый алгоритм RED~\cite{Floyd1993} на основе маршрутизатора
  Cisco. Для построения стенда был выбран программный комплекс GNS3
  (Graphical Network Simulator).
  
  Таким образом, целью исследования является построение на основе GNS3
  виртуального стенда, состоящего из маршрутизатора Cisco, генератора
  трафика и получателя. В качестве генератора трафика используется
  D-ITG (Distributed Internet Traffic Generator).

\section{Стохастическая модель модуля управления трафиком RED}

  Данная модель является развитием жидкостной модели модуля
  управления трафиком RED~\cite{korolkova:2007,
    korolkova:2009, korolkova:2010, Misra1999, Misra2000}. Для унификации методики
  построения модели использовался метод стохастизации одношаговых
  процессов~\cite{kulyabov:2012:vestnik:3,
    kulyabov:2014:vestnik-miph:onestep, ef-kor-gev-kul-sev:vestnik-miph:2014-3}.

  Применяя метод построения одношаговых процессов
  строим стохастическую модель модуля RED, которая содержит два
  основных элемента~--- источник и получатель. В качестве получателя
  рассматривается очередь. Источник отправляет пакеты, получатель
  обрабатывает и посылает подтверждение о принятии пакета.  Строим
  модель исходя из предположения, что источник и получатель
  взаимодействуют по управлению. Таким образом получаем два одномерных
  уравнения, одно из них описывает окно TCP, а второе~--- мгновенную
  длину очереди.  Интенсивность отправки пакетов зависит от размера
  окна.

  Подробно построение стохастической модели управляющего модуля
  управления трафиком типа RED описано
  в~\cite{kulyabov:2014:vestnik:red-sdu} и в приложении
  (раздел~\ref{sec:app:construct}). В статье получена следующая
  модель:

\begin{equation}
  \left\{
  \begin{aligned}
    \d W &= \frac{1}{T} \d t - \frac{W}{2} \d N +
    \sqrt{\frac{1}{T} + \frac{W}{2} \frac{\d N}{\d t}} \d V^1, \\
    \d Q & = \left( \frac{W}{T} - C \right) \d Q + \sqrt{\frac{W}{T}
      - C}\d V^2 ,\\
    \frac{\d \Hat{Q}}{\d t} & = w_q C ( Q - \Hat{Q} ).
  \end{aligned}
  \right.
\end{equation}
Здесь $W$~--- среднее значение размера окна, $Q$~--- среднее значение
мгновенной очереди, $\Hat{Q}$~--- экспоненциально взвешенное
скользящее среднее от среднего мгновенного значения окна со значением
параметра $w_q$, $C$~--- интенсивность исходящего потока из очереди,
$\d V^{i}$~--- двумерный винеровский случайный процесс.

\section{Структура именования образов Cisco IOS}

\begin{figure}
  \centering
  \includegraphics[width=0.8\linewidth]{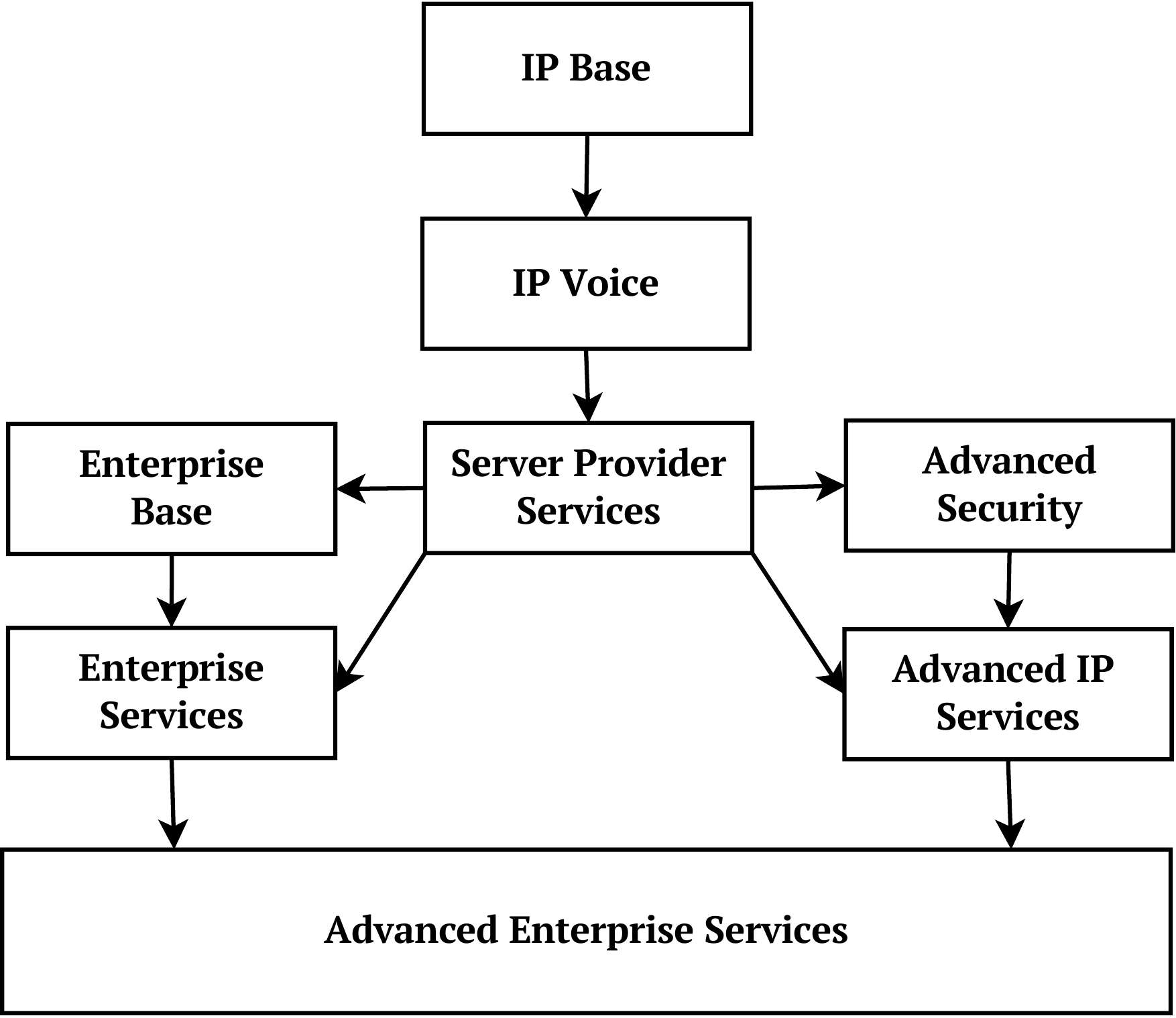}
  \caption{Дерево свойств Cisco IOS}
  \label{fig:ios-tree}
\end{figure}

Для использования образа IOS следует представлять себе, какие
функции поддерживает данный вариант поставки~\cite{cisco:web:cfn}. За это отвечает
параметр <<Feature set>>~(схема~\ref{fig:ios-tree}):

\begin{itemize}
\item IP Base --- начальный уровень функциональности, включается во
  все другие наборы возможностей. Обеспечивает базовый роутинг, то есть
  статические маршруты, RIP, OSPF, EIGRP, только на IPv4. Включает
  VLAN (802.1Q и ISL), которые ранее были доступны только в наборе IP
  Plus. Также включает NAT.
\item IP Services (для коммутаторов 3-го уровня) --- протоколы динамической
  маршрутизации, NAT, IP SLA.
\item Advanced IP Services --- добавляется поддержка IPv6.
\item IP Voice --- добавляет функциональность VoIP и VoFR.
\item Advanced Security --- добавляется IOS/Firewall, IDS, SCTP, SSH и
  IPSec (DES, 3DES и AES).
\item Service Provider Services --- добавляется IPv6, Netflow, SSH,
  BGP, ATM и VoATM.
\item Enterprise Base --- добавляется поддержка унаследованных L3
  протоколов, таких как IPX и AppleTalk. Также включаются IBM features
  типа DLSw+, STUN/BSTUN и RSRB.
\end{itemize}

Начиная с версии IOS 12.3T Cisco использует новую схему именования
образов~(таб.~\ref{tab:ios-feature-new}). Однако данный метод именования не
покрывает всех тонкостей комплектования образа, поэтому до сих пор
используются элементы старой схемы именования~(таб.~\ref{tab:ios-feature-old}).

\begin{table}
\caption{Новая структура именования набора возможностей}
\label{tab:ios-feature-new}
\def\tabIosFeatureNewI{Код}
\def\tabIosFeatureNewII{Набор возможностей}
\begin{center}
\pgfplotstabletypeset[col sep=&,string type,column type=l,
columns={code,feature},
columns/code/.style={column name=\tabIosFeatureNewI},
columns/feature/.style={column name=\tabIosFeatureNewII,column type=p{0.8\linewidth}},
every head row/.style={before row=\toprule,after row=\midrule},
every last row/.style={after row=\bottomrule},
]{
code &  feature
Base &  Entry level image (IP Base, Enterprise Base)
Services &      Addition of IP Telephony Service, MPLS, Voice over IP, Voice over Frame Relay and ATM (Included in SP Services, Enterprise Services)
Advanced &      Addition of VPN, Cisco IOS Firewall, 3DES encryption, SSH, Cisco IOS IPSec and Intrusion Detection Systems (IDS) (Advanced Security, Advanced IP Services)
Enterprise &    Addition of multi-protocols, including IBM, IPX, AppleTalk (Enterprise Base, Enterprise Services)
}
\end{center}
\end{table}

\begin{table}
\caption{Старая структура именования набора возможностей}
\label{tab:ios-feature-old}
\begin{center}
\def\tabIosFeatureOldI{Код}
\def\tabIosFeatureOldII{Набор возможностей}
\pgfplotstabletypeset[col sep=&,string type,column type=l,
columns={code,feature},
columns/code/.style={column name=\tabIosFeatureOldI},
columns/feature/.style={column name=\tabIosFeatureOldII,column type=p{0.8\linewidth}},
every head row/.style={before row=\toprule,after row=\midrule},
every last row/.style={after row=\bottomrule},
]{
code &  feature
I &     IP
Y &     IP on 1700 Series Platforms
S &     IP Plus
S6 &    IP Plus – No ATM
S7 &    IP Plus – No Voice
J &     Enterprise
O &     IOS Firewall/Intrusion Detection
K &     Cryptorgaphy/IPSEC/SSH
K8 &    56Bit DES Encryption (Weak Cryptography)
K9 &    3DES/AES Encryption (Strong Cryptography)
X &     H323
G &     Services Selection Gateway (SSG)
C &     Remote Access Server or Packet Data Serving Node (PDSN)
B &     Apple Talk
N &     Novel IP/IPX
V &     Vox
R &     IBM
U &     Unlawful Intercept
P &     Service Provider
Telco & Telecommunications Feature Set
Boot &  Boot Image (Used on high end routers/switches)
}
\end{center}
\end{table}

По невыясненным причинам не все образы IOS работоспособны в GNS3. Нами
опробовано несколько образов. Далее мы приводим краткий список
работающих и неработающих образов.

Работоспособные образы:
\begin{itemize}
\item \verb|C1700-adventerprisek9-mz.124-8|.
\item \verb|C1710-bk9no3r2sy-mz.124-23|
\item \verb|C1720-l2sy-mz.121-11|
\item \verb|C2600-adventerprisek9_sna-mz.124-25b|
\item \verb|C2691-adventerprisek9_sna-mz.124-23|
\item \verb|C3660-jsx-mz.123-4.T.|
\item \verb|C3745-adventerprisek9_sna-mz.124-15.T14.|
\item \verb|C7200-adventerprisek9_sna-mz.152-4.m4|
\end{itemize}
Неработоспособные образы:
\begin{itemize}
\item \verb|C2600-adventerprisek9-sna-mz.124-23.|
\item \verb|C3745-adventerprisek9_sna-mz.124-15.T14|
\item \verb|C3745-adventerprisek9_ivs-mz.124-15.T14|
\item \verb|C3745-adventerprisek9_ivs-mz.124-15.T8|
\end{itemize}

\section{Установка и настройка GNS3}

  GNS3~--- это комплекс, который позволяет смоделировать виртуальную
  сеть из маршрутизаторов и виртуальных
  машин~\cite{Welsh:2013:GNS3}. Работает практически на всех
  платформах. Фактически это графический интерфейс для разных
  виртуальных машин. Для эмуляции устройств Cisco используется
  эмулятор dynamips. Кроме того, можно использовать такие эмуляторы,
  как VirtualBox и Qemu. Последний особенно удобен при использовании с
  системой KVM, позволяющей использовать аппаратную реализацию
  процессора.

  Графический интерфейс позволяет легко коммутировать разные
  виртуальные машины. Кроме того, есть возможность соединения
  проектируемой топологии с реальной сетью. Использование WireShark
  позволяет провести мониторинг трафика внутри проектируемой
  топологии.

\begin{figure}
   \centering
   \includegraphics[width=0.8\linewidth]{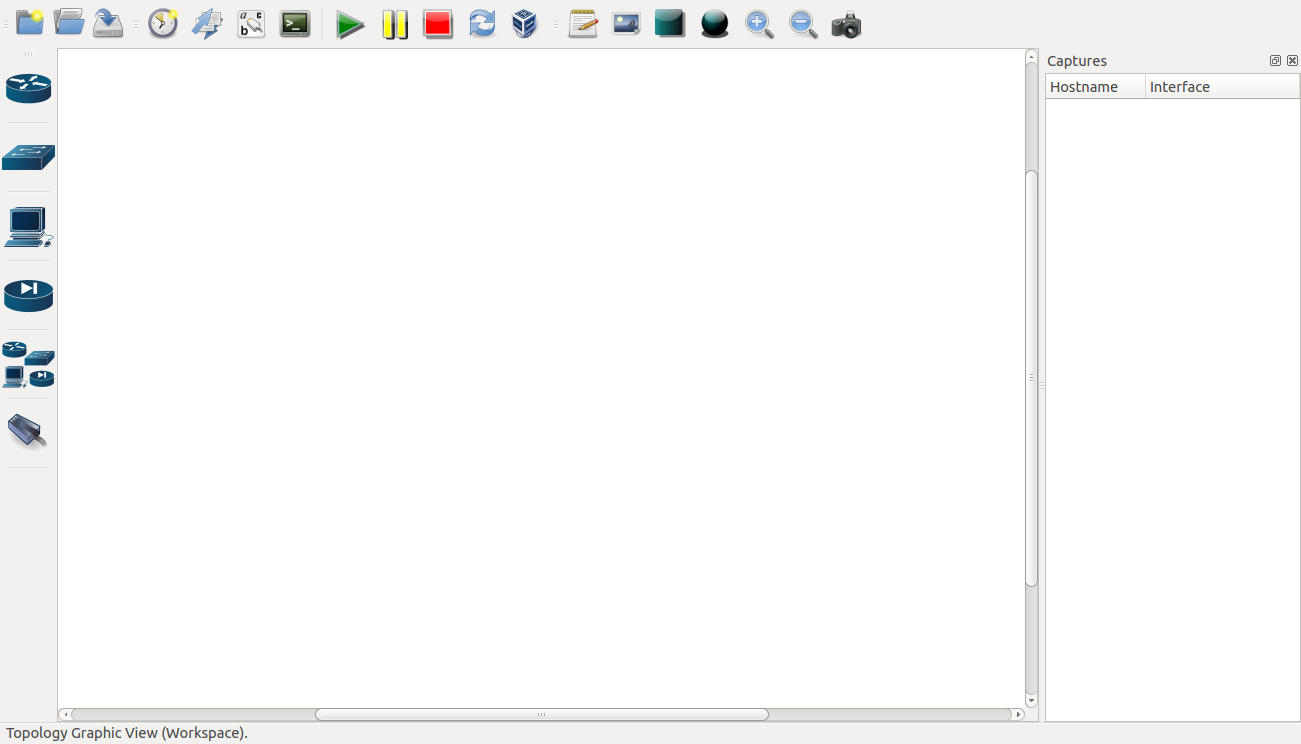}
  \caption{Интерфейс GNS3}
\label{fig:GNS3}
\end{figure}

  Для работы с GNS3 необходимо установить Dynamips, VirtualBox и/или
  QEMU, xdotool, Wireshark.  Чтобы установить вышеуказанное
  программное обеспечение для операционных систем семейства Linux (в
  нашем случае GNS3 устанавливался на Ubuntu 14.04) в консоли
  прописываем следующие команды:

\noindent \verb|sudo apt-get install dynamips|

\noindent \verb|sudo apt-get install qemu|

\noindent \verb|sudo apt-get install virtualbox|

\noindent \verb|sudo apt-get install xdotool|

\noindent \verb|sudo apt-get install wireshark|

После этого устанавливаем сам GNS3, аналогичным образом прописывая
команду в терминале:
\begin{verbatim}
sudo apt-get install gns3
\end{verbatim}

Далее запускаем GNS3. Перед нами открывается интерфейс GNS3
(см. рис.~\ref{fig:GNS3}).

 Вверху расположено контекстное меню, 
 с левой стороны оборудование на выбор, 
 снизу~--- консольное окно программы, справа~--- меню управления 
сетью. Приступая к работе, необходимо провести предварительную
настройку GNS3. 

\begin{figure}
  \centering
  \includegraphics[width=0.8\linewidth]{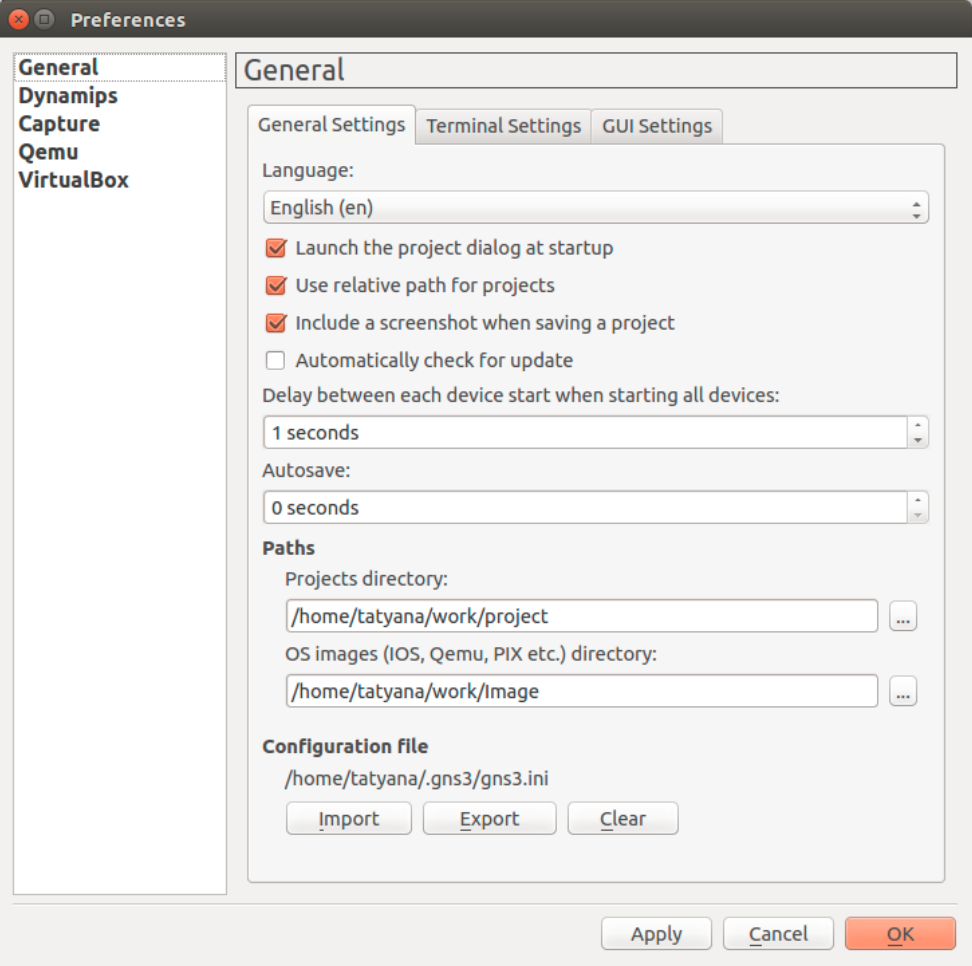}
  \caption{Общие настройки GNS3}
\label{fig:Preferences:G}
\end{figure}

Для этого в контекстном меню во вкладке \menu{Edit} выбираем пункт
\menu{Preferences} (рис.~\ref{fig:Preferences:G}).

\begin{figure}
  \centering
  \includegraphics[width=0.8\linewidth]{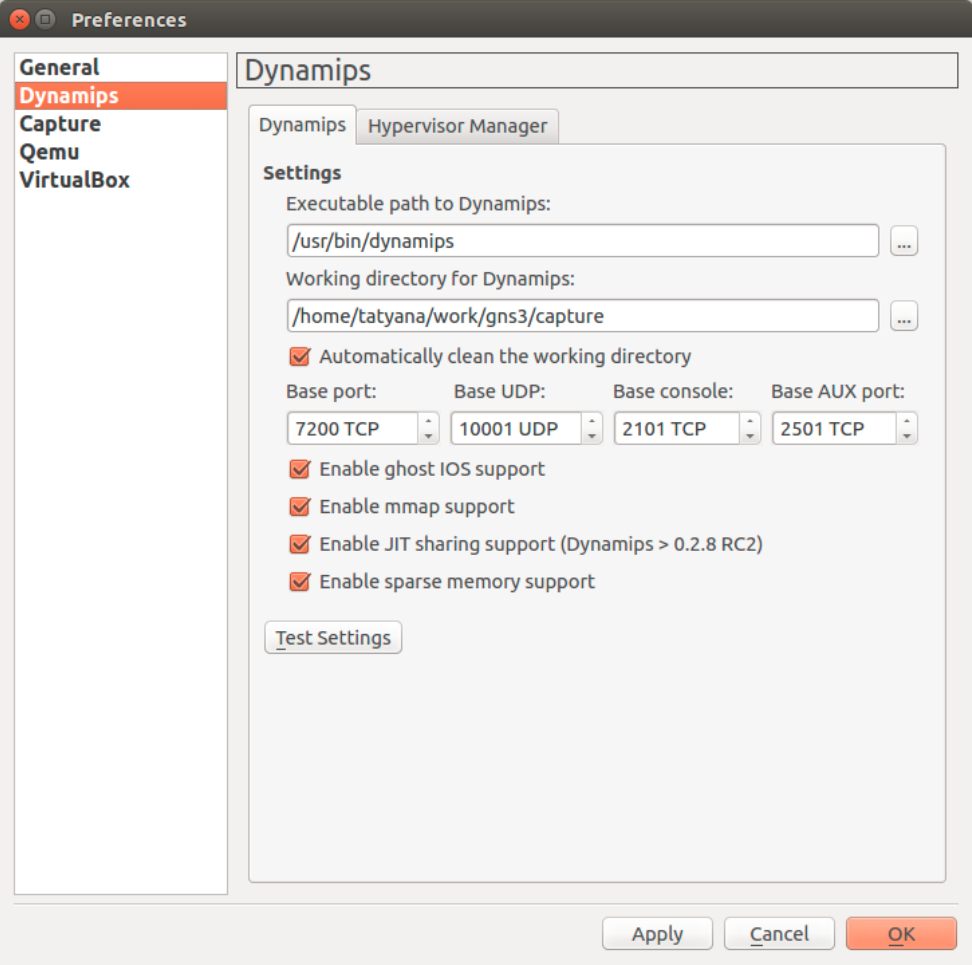}
  \caption{Настройки эмулятора Dynamips}
\label{fig:Preferences:D}
\end{figure}

В подменю \menu{General} есть возможность 
изменения языка.  Здесь прописывается 
путь к папкам, в которых будут хранится проекты и образы 
оборудования.  В \menu{Dynamips} 
прописывается путь к папке, куда установлен
dynamips и путь к папке для захвата файлов
(рис.~\ref{fig:Preferences:D}).

  В подменю \menu{Capture} конфигурируются параметры захвата трафика
  (рис.~\ref{fig:Preferences:C}) (впрочем, на данный момент мы не
  используем возможность захвата трафика).  В строке Default Presents
  выбираем Wireshark Live Traffic Capture (Linux). В следующей строке
  указываем путь к директории, в которой будут храниться данные о
  захвате. И в последней строке прописываем команду для запуска
  Wireshark и чтения файла захвата:
\begin{verbatim}
tail -f -c +0b %c | wireshark -k -i
\end{verbatim}
  
\begin{figure}
  \centering
  \includegraphics[width=0.8\linewidth]{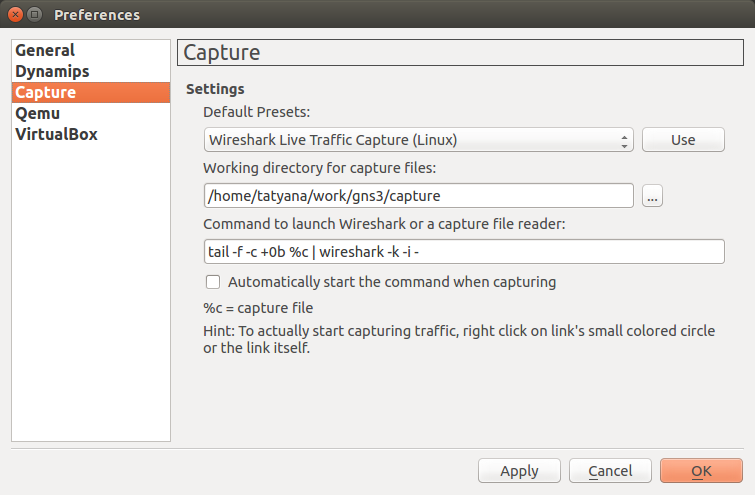}
  \caption{Настройки захвата трафика}
\label{fig:Preferences:C}
\end{figure}

\begin{figure}
   \centering
   \includegraphics[width=0.8\linewidth]{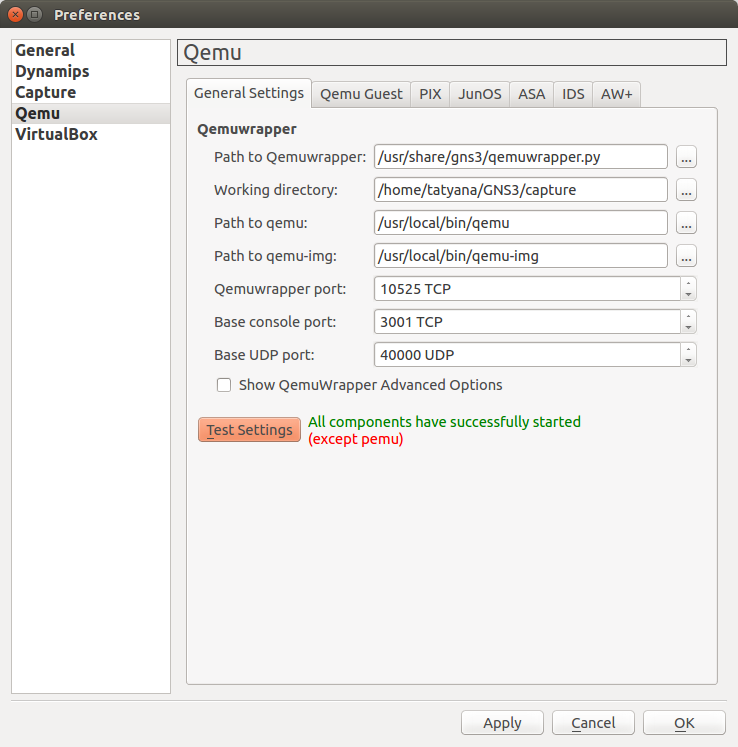}
   \caption{Preferences GNS3 - Qemu}
   \label{fig:Preferences:Qemu:g}
\end{figure}

  В разделе \menu{Qemu} во вкладке \menu{General Settings} прописываем
  путь к Qemuwrapper (этот файл входит в поставку GNS3). Указываем
  директорию для захвата. В строке \menu{Path to qemu} прописываем
  путь к файлу, который эмулирует работу процессора. В следующей
  строке указываем путь к образу виртуальной машины. Номера портов
  оставляем по умолчанию. Чтобы проверить работу Qemuwrapper нажимаем
  на кнопку \menu{Test Settings}, должна появиться зеленая надпись о
  выполнении теста (рис.~\ref{fig:Preferences:Qemu:g}).

Далее, устанавливаем образ ОС на виртуальную машину. Набор образов
можно взять со страницы \url{http://www.gns3.net/appliances/}.
Мы выбрали Linux Core 4.7.7, поскольку он содержит генератор трафика D-ITG.

\begin{figure}
  \centering
  \includegraphics[width=0.8\linewidth]{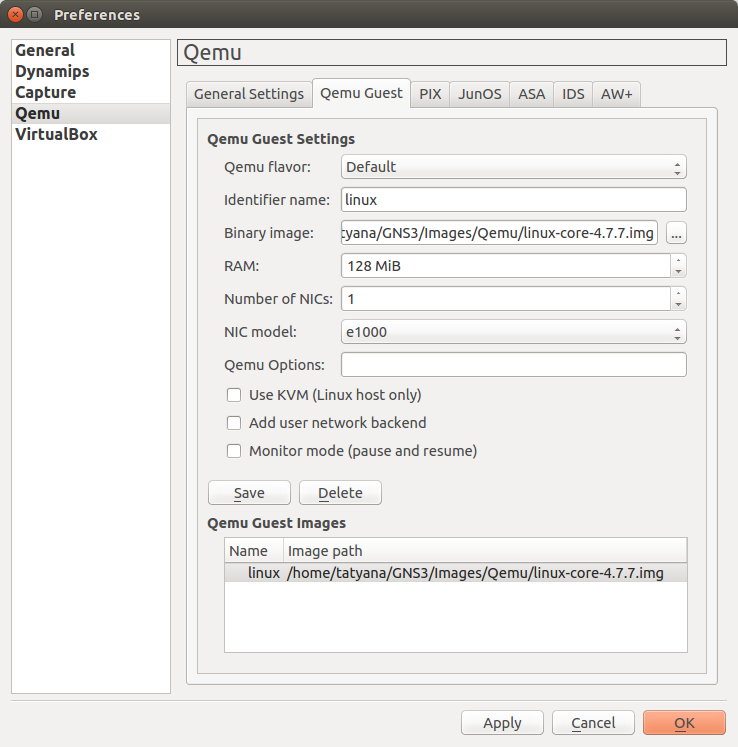}
  \caption{Preferences GNS3}
  \label{fig:Preferences:Qemu:G}
\end{figure}

 В строке \menu{Qemu flavor} выставляем интересующую нас модель процессора,
 указываем имя. После этого прописываем путь к образу ОС.  Нажимаем
 \menu{Save} и \menu{OK}. После этого на панели элементов можно будет
 выбрать \menu{Qemu guest} (рис.~\ref{fig:Preferences:Qemu:G}). 

\begin{figure}
   \centering
   \includegraphics[width=0.8\linewidth]{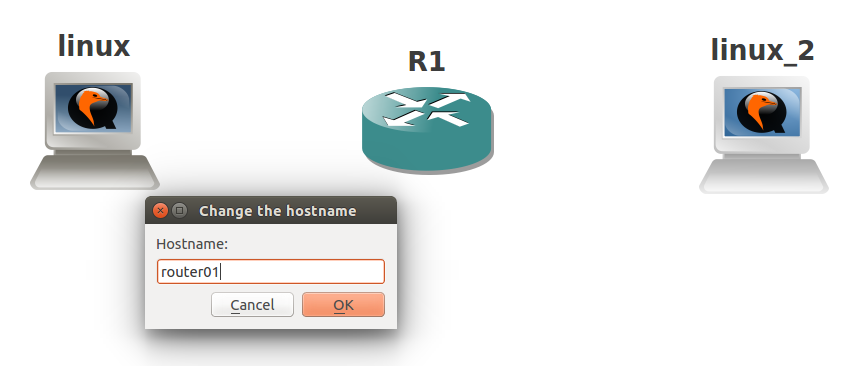}
   \caption{Virtual Network}
   \label{fig:Stand}
\end{figure}

Теперь можно собирать стенд. Для этого переносим из меню выбора
 устройства маршрутизатор и виртуальную машину \menu{Qemu guest},
 переименовываем (\menu{правая кнопка мыши > change the hostname})
 (рис.~\ref{fig:Stand}).

\begin{figure}
   \centering
   \includegraphics[width=0.8\linewidth]{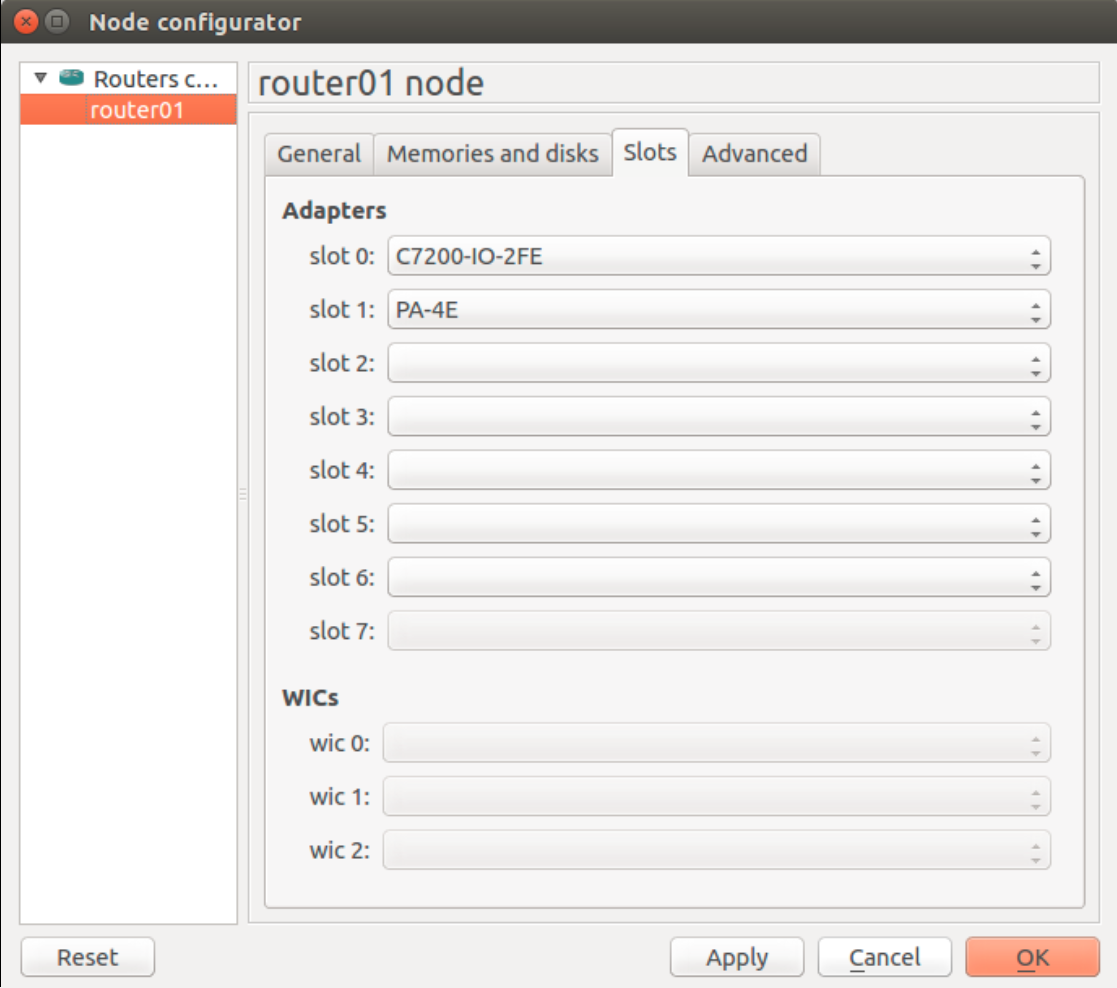}
   \caption{Router configurator}
   \label{fig:Node:R1}
 \end{figure}

На данном стенде, host01~--- источник пакетов; host02~--- получатель. 
Настроим слоты для маршрутизатора. Для 
этого нажимаем правой кнопкой мыши  на 
router01 и выбираем 
\menu{Configure > router01 > Slots}. Для slot0 
выбираем FastEthernet (рис.~\ref{fig:Node:R1}).

Теперь создаем соединение между маршрутизатором и виртуальной машиной. 
Правой кнопкой мыши на host01 выбираем 
\menu{add link > FastEthernet > e0} и соединяем с router01, 
выбирая f0/0. Аналогично этому создаём 
соединение между host02 и router01, только 
теперь выбираем тип соединения Ethernet (скорость 10 Mbps). 

\begin{figure}
  \centering
  \includegraphics[width=0.8\linewidth]{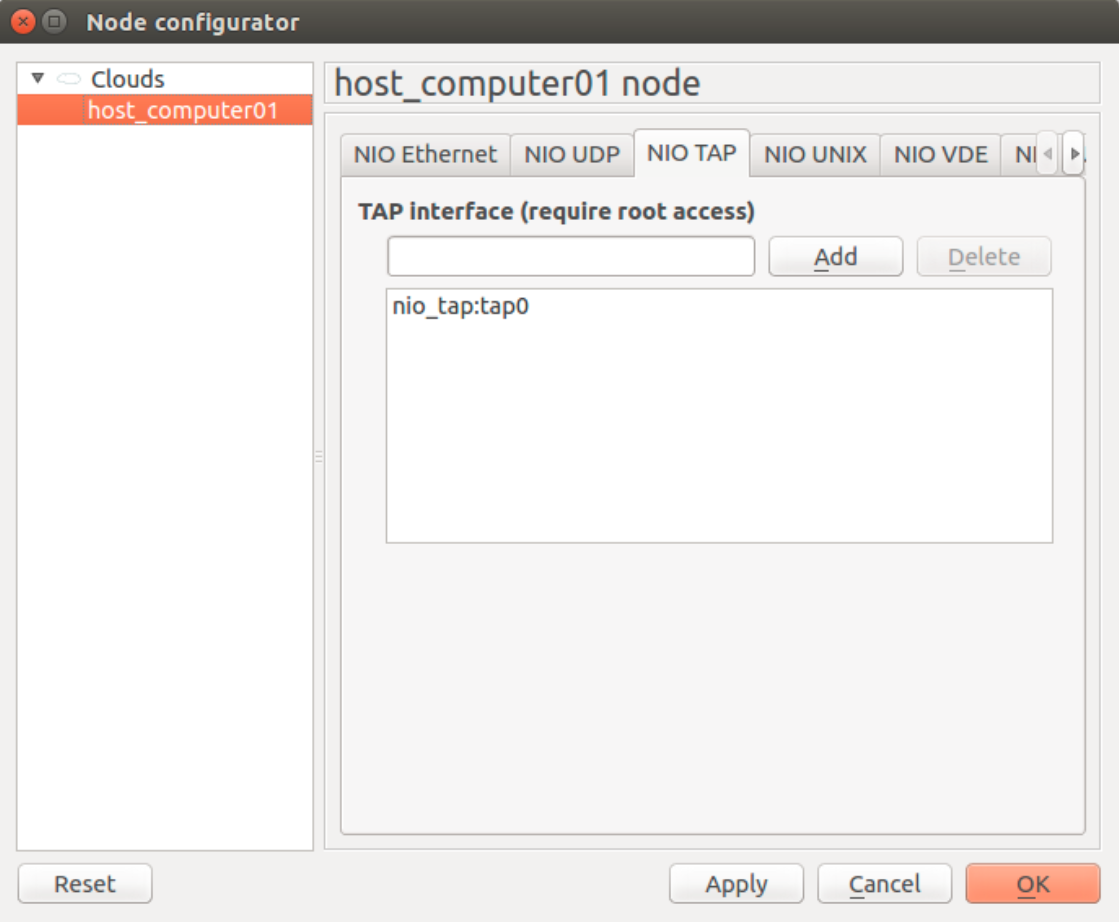}
  \caption{Cloud interface configurator}
  \label{fig:NIO:TAP}
\end{figure}

Теперь нам нужно соединить виртуальные машины с хостовым компьютером
(для получения и обработки логов). Соединение будем осуществлять через
TUN/TAP интерфейс (TAP симулирует Ethernet устройство и работает на
канальном уровне модели OSI, оперируя кадрами Ethernet и используется
для создания моста).
Для этого перетаскиваем на рабочую область облако и 
конфигурируем его: \menu{Configure > NIO TAP}. 
Вписываем название tap0 и нажимаем \menu{add > apply > ok}
(рис.~\ref{fig:NIO:TAP}).  

\begin{figure}
  \centering
  \includegraphics[width=0.8\linewidth]{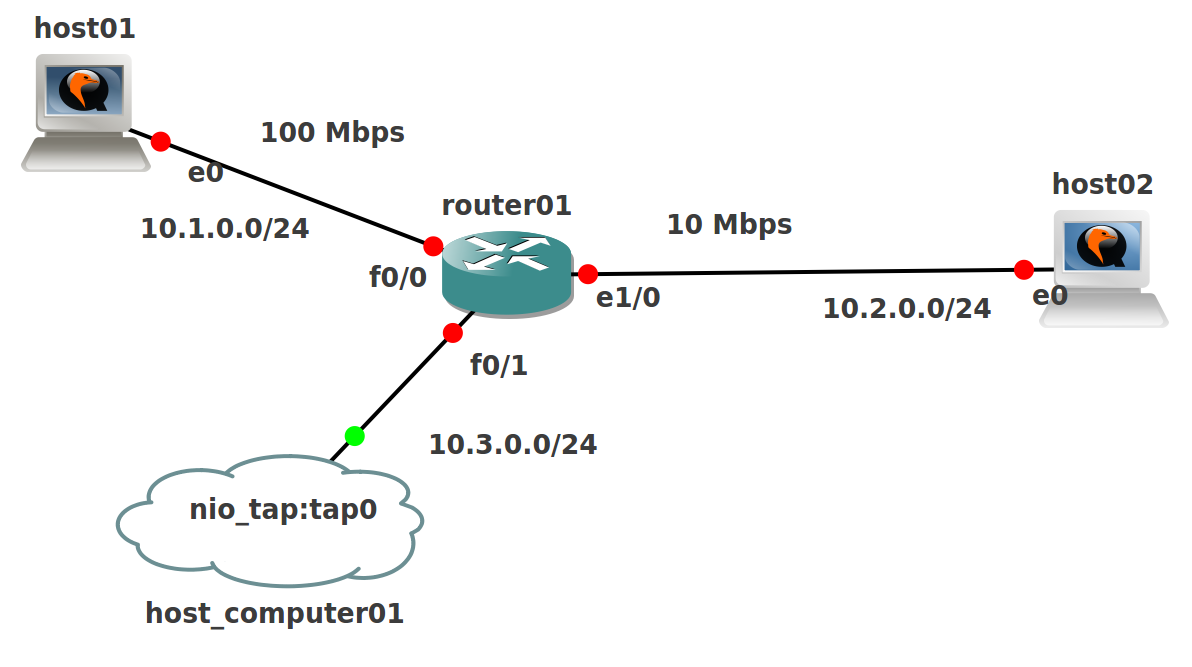}
  \caption{Virtual Network}
  \label{fig:VN}
\end{figure}

Создаём соединение с router01, выбирая тип 
соединения FastEthernet. Получаем готовую топологию (рис.~\ref{fig:VN}).

Далее необходимо произвести конфигурацию устройств.

\begin{figure}
  \centering
  \includegraphics[width=0.8\linewidth]{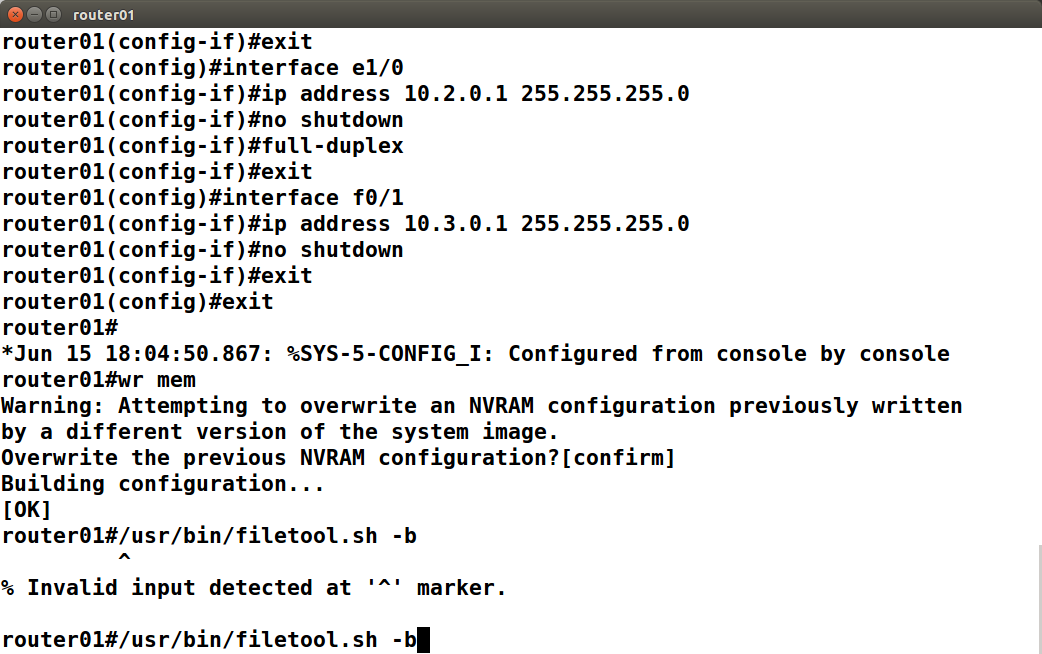}
  \caption{Router CLI configuration}
  \label{fig:router}
\end{figure}

Запускаем router01 (\menu{правая кнопка > start}). Открываем консоль для него 
(\menu{правая кнопка > Console}) и вводим следующие команды (рис.~\ref{fig:router}):
\begin{verbatim}
Router>enable
Router#configure terminal
router01(config)#hostname router01
router01(config)#interface f0/0
router01(config-if)#ip address 10.1.0.1 
  255.255.255.0
router01(config-if)#no shutdown
router01(config-if)#exit
router01(config)#interface e1/0
router01(config-if)#ip address 10.2.0.1 
  255.255.255.0
router01(config-if)#no shutdown
router01(config-if)#full-duplex
router01(config-if)#exit
router01(config)#interface f0/1
router01(config-if)#ip address 10.3.0.1 
  255.255.255.0
router01(config-if)#no shutdown
router01(config-if)#exit
router01(config)#exit
router01#write memory
\end{verbatim}

Теперь открываем консоль для host01. В качестве имени пользователя задаём \verb|tc|, пароль
пустой. Переходим в режим суперпользователя: \verb|sudo su|. 
Редактируем параметры host01 (для начала редактирования в редакторе vi вводим \verb|i|; 
чтобы сохранить изменения и выйти из редактора нажимаем кнопку esc, затем 
вводим \verb|:wq| и нажимаете кнопку \keys{Enter}). Придётся отредактировать
два файла: \verb|/opt/bootsync.sh| и \verb|/opt/bootlocal.sh|:
\begin{verbatim}
vi /opt/bootlocal.sh
\end{verbatim}

В конце файла прописываем следующее:
\begin{verbatim}
ifconfig eth0 10.1.0.10 netmask 255.255.255.0 up
route add default gw 10.1.0.1
\end{verbatim}

  После сохранения изменений в файле выходим из него и выполняем
  команду для сохранения конфигурации (Linux Core использует систему
  конфигурации Tinycore):
\begin{verbatim}
/usr/bin/filetool.sh -b
\end{verbatim}

Аналогично конфигурируется host02, изменяются только ip адреса (сеть 10.2.0.0/24).

Теперь сконфигурим хостовый компьютер, подняв на нём TAP-интерфейс.
В командном окне в режиме суперпользователя создаем интерфейс для
пользователя:
\begin{verbatim}
tunctl -u имя_пользователя -t tap0
\end{verbatim}

Устанавливаем адрес (рис.~\ref{fig:Tap0:1}):
\begin{verbatim}
ifconfig tap0 10.3.0.10 netmask 255.255.255.0 up
\end{verbatim}

\begin{figure}
  \centering
  \includegraphics[width=0.8\linewidth]{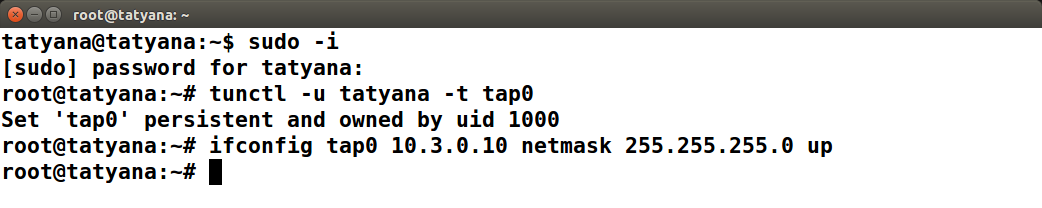}
  \caption{Address configuration on tap0 interface}
  \label{fig:Tap0:1}
\end{figure}

Далее производятся настройка фильтрации и маршрутизация (рис.~\ref{fig:Tap0}):
\begin{verbatim}
iptables -I INPUT 1 -i tap0 -j ACCEPT
route add -net 10.1.0.0/24 dev tap0
route add -net 10.2.0.0/24 dev tap0
\end{verbatim}

\begin{figure}
  \centering
  \includegraphics[width=0.8\linewidth]{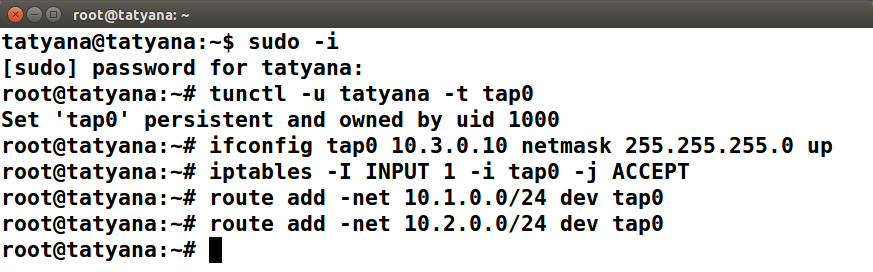}
  \caption{Routing configuration on tap0 interface}
  \label{fig:Tap0}
\end{figure}

\section{Генератор трафика D-ITG}

Теперь можно генерировать трафик и снимать показания. Для генерации
трафика мы используем D-ITG.

D-ITG позволяет получить оценки основных показателей качества
обслуживания (средняя задержка передачи пакетов, вариация задержки
(джиттер), коэффициент потерь пакетов, производительность) с высокой
степенью достоверности. В зависимости от требований
в ходе проведения лабораторного эксперимента можно изменять
следующие параметры:

\begin{itemize}
\item количество потоков, передаваемых между конечными станциями;
\item продолжительность генерирования трафика;
\item интенсивность каждого отдельного потока (пак/с или бит/с);
\item длина пакетов трафика или их закон распределения;
\item вид и параметры закона распределения временного интервала между
соседними пакетами;
\item тип протокола транспортного уровня: TCP, UDP, DCCP, SCTP;
\item тип трафика (имитация потока, создаваемого определенным протоколом
прикладного уровня): VoIP, IPTV, Telnet, DNS, 
игровой трафик (Counter Strike, Quake3). 
\end{itemize}

В рамках D-ITG управление экспериментом осуществляется при помощи
командной строки, и необходимый набор параметров для генерирования
трафика задается путем вызова программы \verb|ITGSend| с использованием
ключей.

\begin{figure}
   \centering
   \includegraphics[width=0.8\linewidth]{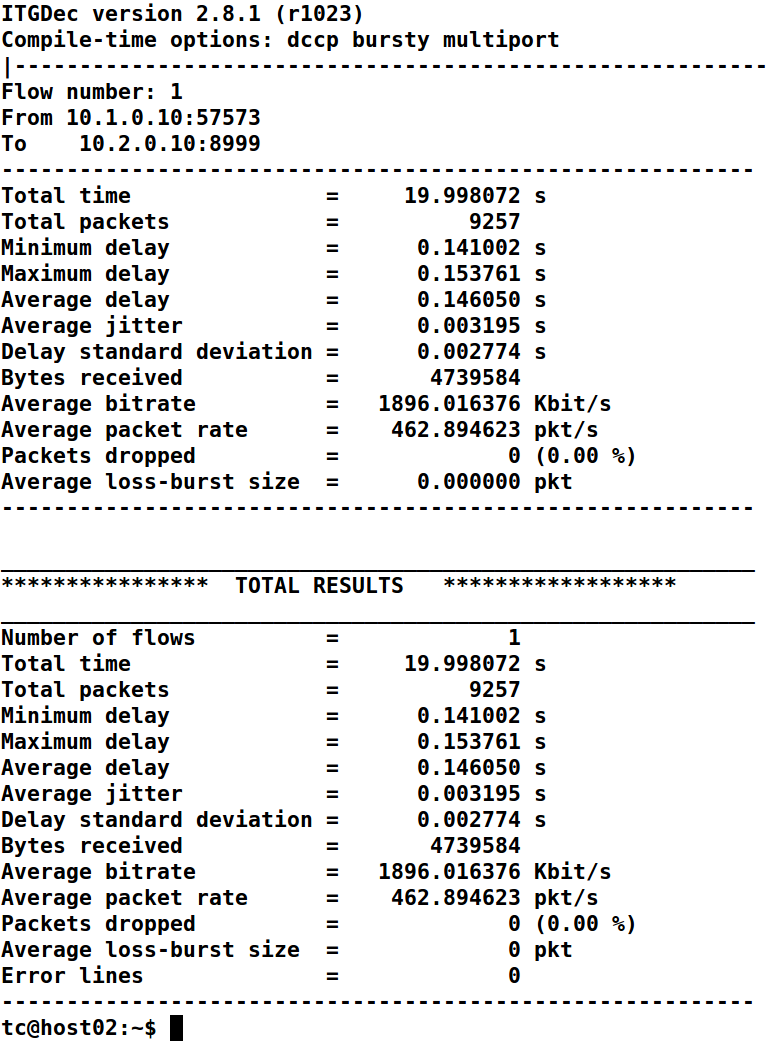}
  \caption{Статистика входного потока. UDP}
\label{fig:UDP}
\end{figure}

Рассмотрим пример генерации трафика по протоколам UDP и TCP. 
В консоли host02 выполняем команду на прослушивание канала:
\begin{verbatim}
ITGRecv
\end{verbatim}

В консоли host01 запустим генерацию трафика:
\begin{verbatim}
ITGSend -a 10.2.0.10 -T UDP -C 10000 -c 500 \
  -t 20000 -x recv_log_file
\end{verbatim}

Здесь мы передаем файлы получателю с адресом 10.2.0.10 по протоколу
UDP, скорость передачи 10000 пакетов в секунду, размер пакета 500
байт, время соединения 20000 мс.  Вся
информация по отправке данных записывается на стороне получателя в
файл с названием
\verb|recv_log_files|. После того как передача данных была выполнена,
останавливаем прослушивание в консоли host02 
(нажимая \keys{Ctrl + C}), затем в консоли host02 выполняем команду на
декодирование лог-файла: 
\begin{verbatim}
ITGDec recv_log_file
\end{verbatim}

При помощи данной команды на экран 
выводится таблица со значениями параметров принятого потока (рис.~\ref{fig:UDP}).

\begin{figure}
  \centering
  \includegraphics[width=0.8\linewidth]{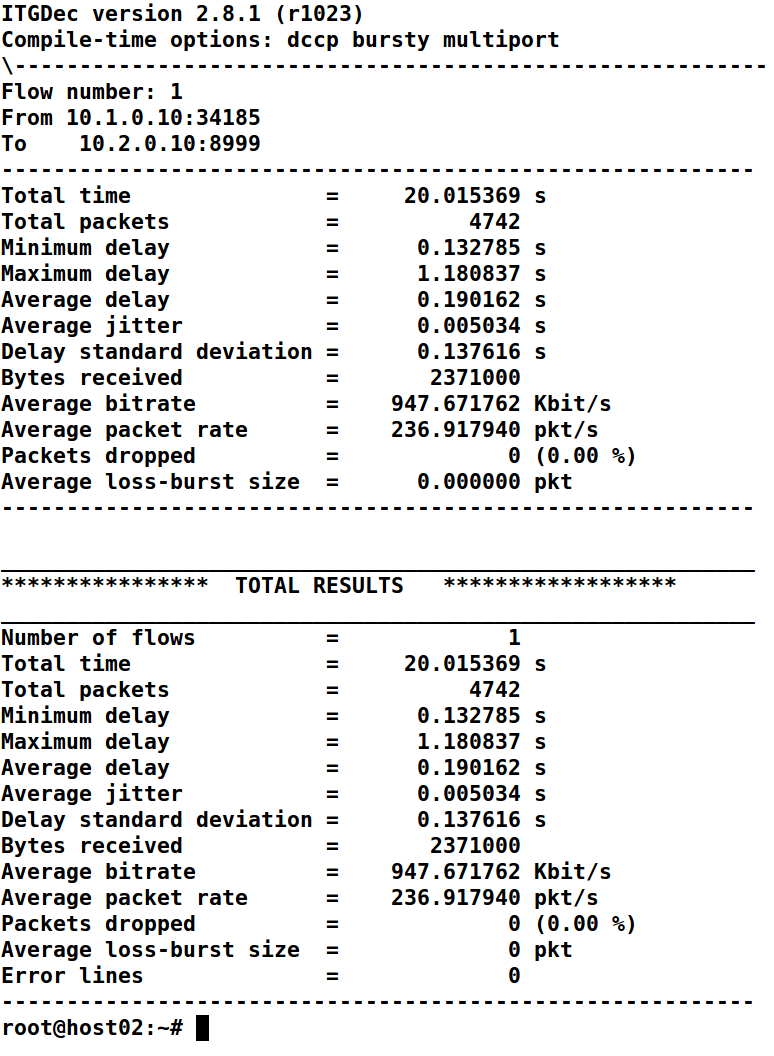}
  \caption{Статистика входного потока. TCP}
  \label{fig:TCP}
\end{figure}

Аналогично выполняем для TCP и получаем следующее (рис.~\ref{fig:TCP}).

\begin{figure}
  \centering
  \includegraphics[width=0.8\linewidth]{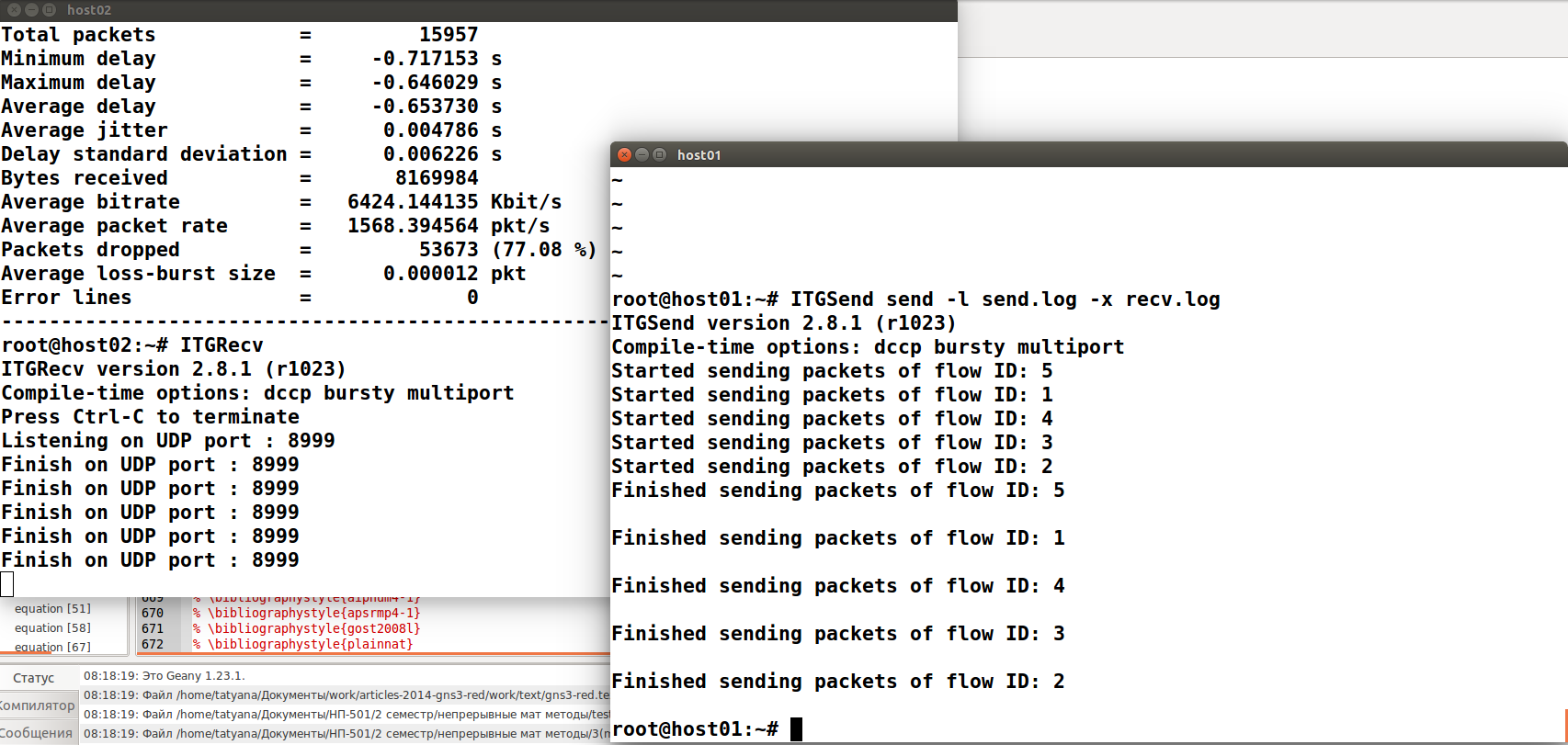}
  \caption{Дешифровка лога}
\label{fig:Command:UDP}
\end{figure}

Для многопотоковой передачи трафика в командном окне host01 в файле
\verb|send| построчно описываем каждый поток. Например, таким образом: 

\begin{verbatim}
cat > send <<EOF
-a 10.2.0.10  -C 1000 -c 512 -T UDP
-a 10.2.0.10  -C 2000 -c 512 -T UDP
-a 10.2.0.10  -C 3000 -c 512 -T UDP
-a 10.2.0.10  -C 4000 -c 512 -T UDP
-a 10.2.0.10  -C 5000 -c 512 -T UDP
EOF
\end{verbatim}

Теперь запускаем передачу трафика на host01. 
Данные о передаче трафика с host01 на host02 
записываются в файлы send.log и recv.log соответственно:
\begin{verbatim}
ITGSend send -l send.log -x recv.log
\end{verbatim}

Выведем на экран отчет о передаче 
потоков. Для этого в консоли host02 напишем команду 
\begin{verbatim}
ITGDec send.log
\end{verbatim}
для дешифровки отправляемого трафика (рис.~\ref{fig:Command:UDP}) и
команду
\begin{verbatim}
ITGDec recv.log
\end{verbatim}
для дешифровки получаемого трафика.

Таким образом получаем данные о каждом потоке, а 
также значение параметров исходящего
и входящего потоков.

\section{Визуализация результатов}

По результатам эксперимента можно построить
графики для разных характеристик, как то: 
\begin{itemize}
\item задержка передачи пакетов (delay);
\item вариация задержки (jitter);
\item интенсивность (bitrate);
\item коэффициент потерь пакетов (packet loss). 
\end{itemize}

\begin{figure}
  \centering
  \includegraphics[width=0.8\linewidth]{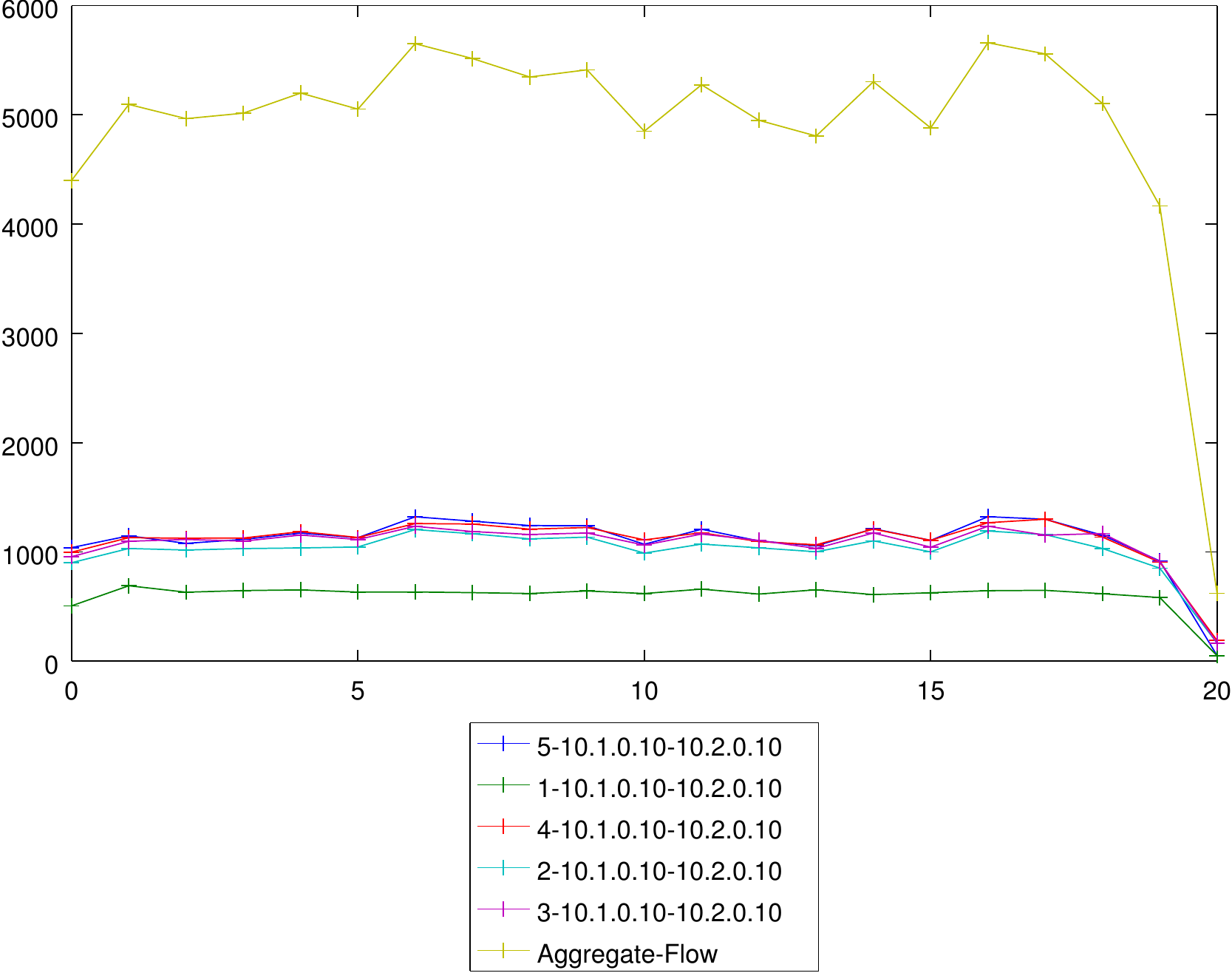}
  \caption{Потеря пакетов (UDP)}
\label{fig:Command:UDP:Packetloss}
\end{figure}

Рассмотрим пример построения графиков.  Предположим, что у нас есть 5
потоков, которые передаются по протоколу TCP.  Скорость передачи 1
потока~--- 1000 пакетов в секунду, 2 потока~--- 2000 пакетов в секунду,
3 потока~--- 3000 пакетов в секунду, 4 потока~--- 4000 пакетов в
секунду, 5 потока~--- 5000 пакетов в секунду.  Размеры пакетов у всех
потоков одинаковы~--- 512 байт, а время соединения 20000 мс.
Для демонстрации потери пакетов будем использовать протокол UDP
(рис.~\ref{fig:Command:UDP:Packetloss}). 

\begin{figure}
   \centering
   \includegraphics[width=0.8\linewidth]{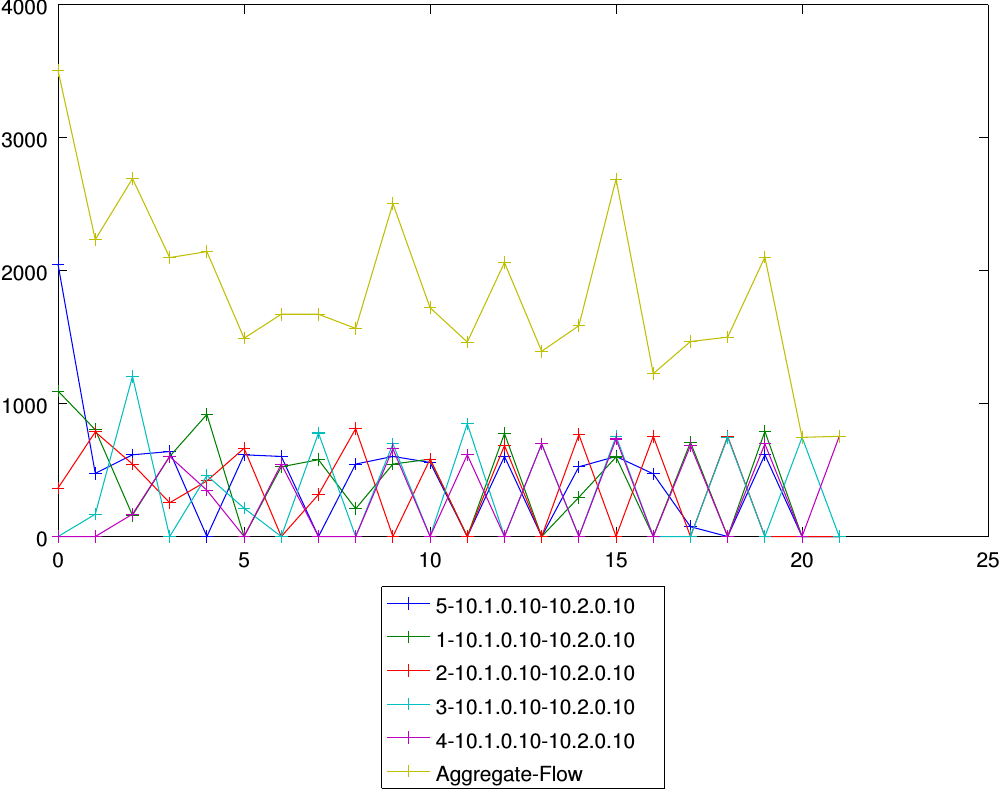}
  \caption{Интенсивность поступающего потока}
 \label{fig:Command:TCP:Bitrate:host01}
\end{figure}

\begin{figure}
  \centering
  \includegraphics[width=0.8\linewidth]{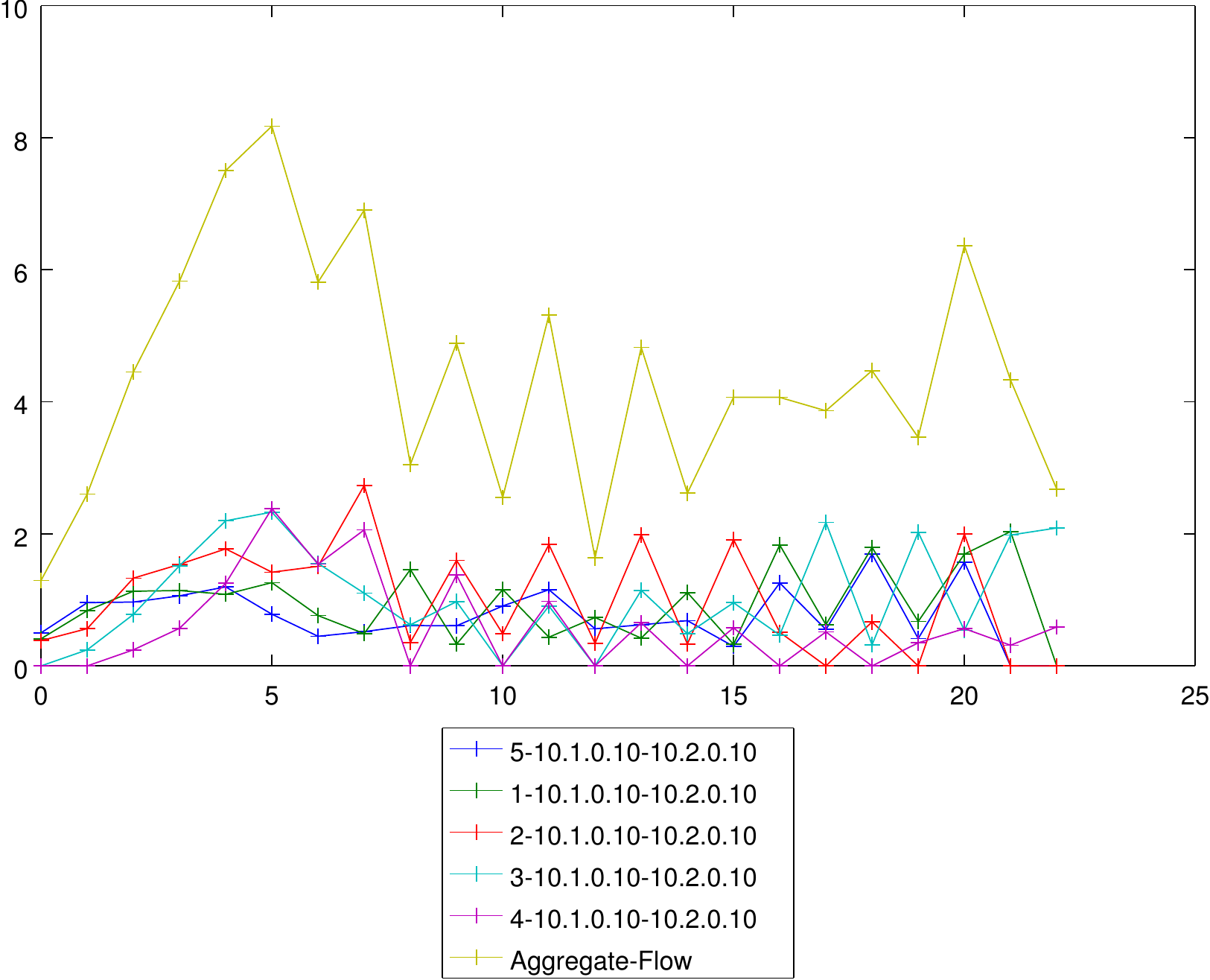}
  \caption{Задержка получения пакетов}
\label{fig:Command:TCP:Delay:host02}
\end{figure}

\begin{figure}
  \centering
  \includegraphics[width=0.8\linewidth]{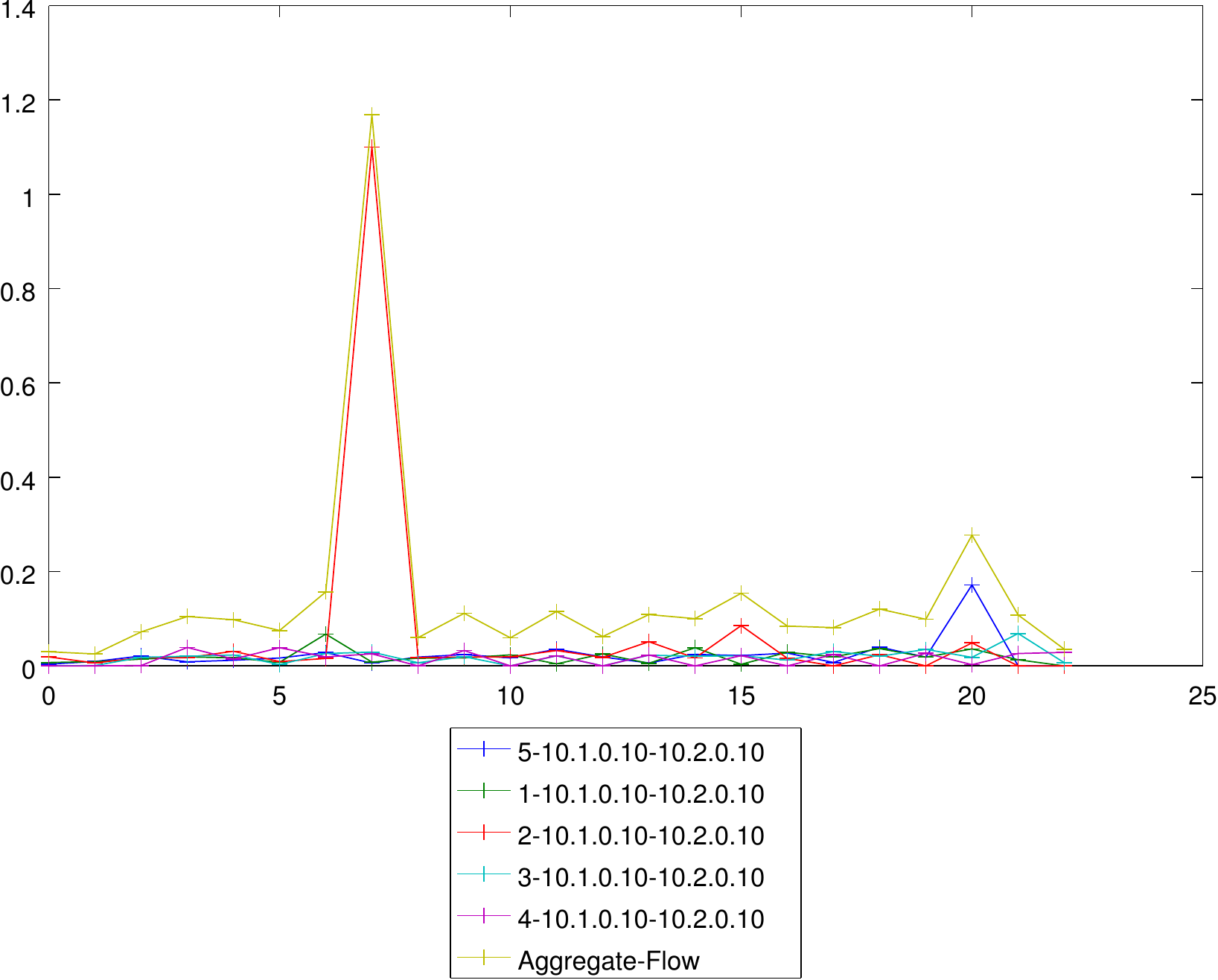}
  \caption{Неравномерность получения пакетов}
\label{fig:Command:TCP:jitter}
\end{figure}

При помощи \verb|ITGplot| построим графики для bitrate, delay и jitter. Для этого
нужно записать полученные значения в отдельные файлы при помощи
следующих команд:
\begin{verbatim}
ITGDec recv.log -b 1000 
ITGDec recv.log -d 1000
ITGDec recv.log -j 1000  
\end{verbatim}

Здесь через каждые 1000 мс в файлы bitrate.dat, jitter.dat, delay.dat
записываются соответственные значения параметров передачи для потоков
(рис.~\ref{fig:Command:TCP:Bitrate:host01},
\ref{fig:Command:TCP:Delay:host02}, \ref{fig:Command:TCP:jitter}).
Построим графики при помощи следующих команд:
\begin{verbatim}
./ITGplot birate.dat
./ITGplot delay.dat
./ITGplot jitter.dat
\end{verbatim}
На полученных графиках верхняя кривая 
отображает общее поведение всех потоков.

\section{Выводы}

Таким образом, мы построили основу стенда для верификации модели
модуля управления трафиком RED. Однако, в данной работе не рассмотрена
настройка самого модуля RED на маршрутизаторе и принципы снятия
показаний с роутера в реальном времени.

\appendix*

\section{Построение стохастической модели модуля управления трафиком RED}
\label{sec:app:construct}

Изменение окна при каждом подтверждении (ACK)
\begin{equation}
  W_{n+1} = W_n + \frac {1}{W_n}.
\end{equation}

Для применения метода стохастизации одношаговых процессов мы
должны перейти к модели с непрерывным временем.
Непрерывное уравнение для окна
$T$~--- время двойного оборота. За это время приходят $W_{n}$
подтверждений. 
\begin{equation}
  \Dot{W} = \frac{1}{T}.
\end{equation}

Дискретное уравнение мгновенной длины очереди:
\begin{equation}
  Q_{n+1} = Q_{n} + W_{n} - C_{n}.
\end{equation}
 
На основе дискретного уравнения 
мгновенной длины очереди выводится 
непрерывное уравнение, которое записывается следующим образом:
\begin{equation}
  \Dot{Q} = \frac{W(t)}{T(t)} - C(t).
\end{equation}

Далее опишем поведение экспоненциально взвешенной скользящей средней
длины очереди, которая представляет собой уравнение связи между источником
и получателем.

Дискретное реккурентное уравнение средней длины очереди 
\begin{equation}
  \Hat{Q}(t_{k} + \delta) = (1 - w_q)\Hat{Q}(t_k) +w_q Q(t_k).
\end{equation}
Из дискретного уравнение выводим непрерывное уравнение средней длины
очереди
\begin{equation}
  \Dot{\Hat{Q}}= - \frac{w_q}{\delta} \Hat{Q} + \frac{w_q}{\delta} Q.
\end{equation}

Чтобы рассмотреть поведение пакетов в системе строим кинетическое
уравнение или как его еще называют уравнение взаимодействия.
Количество пакетов задаётся окном TCP.
\begin{equation}
  \begin{cases}
    0 \xrightarrow{k^1_1} W,\\
    W \xrightarrow{k^1_2} 0.
  \end{cases}
\end{equation}
Первое соотношение описывает появление пакетов в системе, второе~--- вывод пакетов из
системы.

На основании записанных уравнений, применяя метод построения
одношаговых процессов, выводим уравнение Фоккера--Планка:
\begin{multline}
  \frac{\partial w}{\partial t} = 
  - \frac{\partial}{\partial W} 
  \left[ \left( \frac{1}{W} - \frac{W}{2} \frac{\d N}{\d t} \right) w \right]
  + {} \\ {} +
  \frac{1}{2} \frac{\partial^2}{\partial W^2} 
  \left[ \left( \frac{1}{W} + \frac{W}{2} \frac{\d N}{\d t} \right) w \right].
\end{multline}
Из уравнения Фоккера--Планка получаем соответствующее ему уравнение
Ланжевена:
\begin{equation}
  \label{eq:W:langevin}
  \d W = \frac{1}{W} \d t - \frac{W}{2} \d N +
  \sqrt{\frac{1}{W} + \frac{W}{2} \frac{\d N}{\d t}} \d V^1,
\end{equation}
где $\d V^1$~--- винеровский процесс, соответствующий случайному
процессу $W(t)$.

Аналогичным образом описывается поведение длины очереди.
Уравнения взаимодействия для мгновенной длины очереди:
\begin{equation}
  \begin{cases}
    0 \xrightarrow{k^2_1} Q\\
    0\xrightarrow{k^2_2} Q.
  \end{cases}
\end{equation}

Уравнение Фоккера--Планка для мгновенной длины очереди:
\begin{equation}
  \frac{\partial q}{\partial t} = 
  - \frac{\partial}{\partial Q} 
  \left[ \left( \frac{W}{T} - C \right) q \right]
  + \frac{1}{2} \frac{\partial^2}{\partial Q^2} 
  \left[ \left( \frac{W}{T} - C \right) q \right].
\end{equation}

Уравнение Ланжевена для мгновенной длины очереди:
\begin{equation}
  \label{eq:Q:langevin}
  \d Q = \left( \frac{W}{T} - C \right) \d t +
  \sqrt{\frac{W}{T} - C} \d V^2,
\end{equation}
где $\d V^2$~--- винеровский процесс, соответствующий случайному
процессу $Q$.

Согласно полученным уравнениям записываем результирующую систему:
\begin{equation}
  \left\{
  \begin{aligned}
    \d W &= \frac{1}{T} \d t - \frac{W}{2} \d N +
    \sqrt{\frac{1}{T} + \frac{W}{2} \frac{\d N}{\d t}} \d V^1, \\
    \d Q & = \left( \frac{W}{T} - C \right) \d Q + \sqrt{\frac{W}{T}
      - C}\d V^2 ,\\
    \frac{\d \Hat{Q}}{\d t} & = w_q C ( Q - \Hat{Q} ).
  \end{aligned}
  \right.
\end{equation}
Наиболее подробно стохастическая модель управляющего модуля
маршрутизатора типа RED описана
в~\cite{kulyabov:2014:vestnik:red-sdu}.

\bibliographystyle{gost2008l}

\bibliography{bib/gns3-red,bib/gns3-red_ru,bib/gns3-red_en}